\documentclass[pra,twocolumn,showpacs,preprintnumbers,amsmath,amssymb]{revtex4}

\usepackage{graphicx}% Include figure files
\usepackage{dcolumn}% Align table columns on decimal point
\usepackage{bm}% bold math
\usepackage{color}% bold math

\begin{document}

\preprint{APS/123-QED}

%\begin{minipage}{18cm}

\title{Spontaneous Magnetic Ordering in a Ferromagnetic Spinor Dipolar Bose-Einstein Condensate}
\author{Yuki Kawaguchi$^1$}
\author{Hiroki Saito$^2$}
\author{Kazue Kudo$^3$}
\author{Masahito Ueda$^{1,4}$}
\affiliation{
$^1$Department of Physics, University of Tokyo, 7-3-1 Hongo, Bunkyo-ku, Tokyo 113-0033, Japan\\
$^2$Department of Engineering Science, University of Electro-Communications,
1-5-1 Chofugaoka, Chofu-shi, Tokyo 182-8585, Japan\\
$^3$Ochadai Academic Production, Ochanomizu University, 2-1-1 Ohtsuka, Bunkyo-ku, Tokyo 112-8610, Japan\\
$^4$Macroscopic Quantum Control Project, ERATO, JST, Bunkyo-ku, Tokyo 113-8656, Japan
}
\date{\today}

\begin{abstract}
We study the spin dynamics in a spin-1 ferromagnetic Bose-Einstein condensate with magnetic dipole-dipole interaction (MDDI)
based on the Gross-Pitaevskii and Bogoliubov theories.
We find that various magnetic structures such as checkerboards and stripes emerge
in the course of the dynamics due to the combined effects of
spin-exchange interaction, MDDI, quadratic Zeeman and finite-size effects, and non-stationary initial conditions.
However, 
the short-range magnetic order observed by the Berkeley group
 [Phys. Rev. Lett. {\bf 100}, 170403 (2008)] is not fully reproduced in our calculations;
the periodicity of the order differs by a factor of three and the checkerboard pattern eventually dissolves in our numerical simulations.
Possible reasons for the discrepancy are discussed.
\end{abstract}

%03.75.Mn Multicomponent condensates; spinor condensates 
%03.75.Kk Dynamic properties of condensates; collective and hydrodynamic excitations, superfluid flow 
%67.30.he Textures and vortices

%\keywords{keywords}%Use showkeys class option if keyword
\maketitle
%\end{minipage}
%\tableofcontents

\section{Introduction}
%%% Introduction %%%%%%%%%%%%%%%%%%%%%%%%%%%%%%%%%%%%%%%%%%%%%
Experiments on dilute Bose-Einstein condensates (BECs) have exhibited a rich variety of phenomena, most of which have successfully been explained by theory,
including the system
with a strong magnetic dipole-dipole interaction (MDDI)~\cite{Lahaye2008}.
The magnetic crystallization of a spin-1 ferromagnetic $^{87}$Rb condensate recently observed by the Berkeley group~\cite{Vengalattore2008,Vengalattore2009}
provides one of the few anomalies that have so far defied theoretical explanation. 
It has been argued that the MDDI, which is long-ranged and anisotropic, 
plays a pivotal role in the magnetic crystallization~\cite{Vengalattore2008,Vengalattore2009,Lamacraft2008,Sau2009,Cherng2009,Kjall2009,Zhang2009};
however, no satisfactory account of the experiment has been presented.

%%% Berkeley experiment %%%%%%%%%%%%%
The aim of this paper is to clarify what the mean-field and Bogoliubov theories predict under the conditions of the Berkeley experiment~\cite{Vengalattore2008}.
The authors of Ref.~\cite{Vengalattore2008}
have found that a helical spin structure of a quasi-two-dimensional (2D)
condensate spontaneously develops into a short-range crystalline pattern of magnetic domains
in a time scale of a few hundreds of milliseconds.
The typical size of the magnetic pattern is $\lambda_{\rm exp}\simeq 10~\mu{\rm m}$ (the size of the magnetic domain is $\lambda_{\rm exp}/2$), and 
the growth rate of the crystalline pattern increases with decreasing the pitch of the initial spin helix.
On the other hand, when the initial spin configuration is uniform, the short-range magnetic patterns do not emerge over a period of 300 ms.
During the dynamics, the condensate remains fully magnetized and 
the longitudinal magnetization is much smaller than the transverse magnetization.
They also found that the number of spin vortices increases as the short-range magnetic patterns develop.
The nature of the pattern is insensitive to the strength of the quadratic Zeeman energy $q$ in the range of $0.8 < q/h < 4$ Hz.
The same group also reported the emergence of a similar crystalline pattern by cooling initially unmagnetized gases~\cite{Vengalattore2009}.
This result suggests that the system of spin-1 $^{87}$Rb BEC possesses an intrinsic mechanism that stabilizes the crystalline magnetic order with a characteristic size of $\lambda_{\rm exp}$.

The complexity of this system arises from the interplay among three effects related to the magnetism:
the short-range ferromagnetic spin-exchange interaction,
the quadratic Zeeman effect which favors transverse magnetization~\cite{Stenger1998, Murata2007},
and the long-range anisotropic MDDI which induces spatial spin textures~\cite{Yi2006, Kawaguchi2006b, Takahashi2007}.
In addition to them, the initial conditions and noises, the finite-size effect and nonuniform density profile due to the trapping potential
contribute to the dynamics.
In this paper, we take into account all these features of the spinor dipolar BEC and investigate the extent to which we can understand the observed phenomena.
We do find the emergence of magnetic checkerboard patterns due to the inhomogeneity and finite-size effects of the trapping potential;
however, the periodicity differs by at least a factor of three and the checkerboard pattern eventually dissolves in our numerical simulations.
We find that the MDDI induces magnetic patterns that are not checkerboard patterns
but spin-helix or staggered-domain structures depending on the magnitude of the quadratic Zeeman energy.

%% what we have done in this work %%%%%
This paper is organized as follows.
In Sec.~\ref{sec:SpinorDipolarBEC}, we describe a system of the spin-1 spinor dipolar BEC, 
and review the ground-state properties in the absence of the MDDI.
In Sec.~\ref{sec:linearstability}, we investigate the linear stability of the spin-1 spinor dipolar BEC in an infinite quasi-2D system.
We analytically solve the Bogoliubov equation for a uniform spin structure.
The Bogoliubov spectrum for a helical spin structure is numerically obtained.
In Sec.~\ref{sec:sim_pancake}, we discuss the spin dynamics in an oblate trap.
We numerically solve the three-dimensional (3D) Gross-Pitaevskii equation (GPE),
and examine the individual effect of the trapping potential, initial spin helix, and the MDDI.
In Sec.~\ref{sec:compare}, we compare the obtained results with the Berkeley experiment~\cite{Vengalattore2008}.
The possible reasons for the discrepancy between the experimental results and our results are discussed in Sec.~\ref{sec:discussion}.
We provide concluding remarks in Sec.~\ref{sec:conclusions}.

\section{Spin-1 Spinor Dipolar BEC}
\label{sec:SpinorDipolarBEC}
\subsection{Mean-field energy}
%%%% Formalism of a spin-1 BEC + dipole %%%% 
We consider a spin-1 BEC of $N$ atoms confined in an optical trap $U_{\rm trap}({\bm r})$.
The zero-temperature mean-field energy is given by
\begin{align}
E =& \int d{\bm r} \sum_{m=0,\pm1}\Psi_m^*({\bm r})\left[ -\frac{\hbar^2\nabla^2}{2M}+U_{\rm trap}({\bm r})\right]\Psi_m({\bm r})\nonumber\\
   &+ E_{\rm s}+ E_{\rm Z1} + E_{\rm Z2} + E_{\rm dd}
\label{eq:energy0}
\end{align}
where $M$ is the atomic mass,
$\Psi_m({\bm r})$ is the order parameter of the condensate with magnetic sublevel $m=0, \pm 1$,
and $E_{\rm s}, E_{\rm Z1}, E_{\rm Z2}$ and $E_{\rm dd}$ are
the short-range interaction energy, the linear and quadratic Zeeman energy, and the long-range MDDI energy, respectively.
The order parameter is normalized to satisfy $\sum_m\int d{\bm r}|\Psi_m|^2=N$.

% short-range interaction
The short-range interaction energy is given by
\begin{align}
E_{\rm s}=\frac{1}{2} \int d{\bm r} \left[c_0 n^2({\bm r}) + c_1 |{\bm f}({\bm r})|^2\right],
\label{eq:Es}
\end{align}
where
\begin{align}
n({\bm r})=\sum_m|\Psi_m({\bm r})|^2
\end{align}
is the atom-number density and
\begin{align}
{\bm f}({\bm r}) = \sum_{mm'}\Psi_m^*({\bm r}) {\bm F}_{mm'}\Psi_{m'}({\bm r})
\end{align}
is the spin density with ${\bm F}=(F_x,F_y,F_z)$ being the vector of the spin-1 matrices,
and the interaction coefficients in Eq.~\eqref{eq:Es} are given by
\begin{align}
c_0 &= \frac{4\pi\hbar^2}{M} \frac{a_0 + 2a_2}{3},\\
c_1 &= \frac{4\pi\hbar^2}{M} \frac{a_2 - a_0}{3},
\end{align}
with $a_S$ $(S=0,2)$ being the {\it s}-wave scattering length for the scattering channel with total spin $S$.

% linear Zeeman effect
In the presence of an external magnetic field ${\bm B}\equiv B\hat{\bm e}_B$,
the linear Zeeman energy is given by
\begin{align}
E_{\rm Z1} &= \int d{\bm r} 
\sum_{mm'}  \Psi_m^*({\bm r})\left(\hbar\omega_{\rm L}\hat{\bm e}_B\cdot {\bm F}\right)_{mm'}\Psi_{m'}({\bm r}),
\end{align}
where $\omega_{\rm L}=g_F\mu_{\rm B}B/\hbar$ with $g_F$ being the hyperfine g-factor and $\mu_{\rm B}$ the Bohr magneton.
In this paper, we take the spin quantization axis $z$ along the external magnetic field, i.e., $\hat{\bm e}_B=\hat{z}$.
Then, the linear Zeeman energy is expressed as
\begin{align}
E_{\rm Z1} &= \int d{\bm r}  \sum_{m}  \hbar\omega_{\rm L} m |\Psi_m({\bm r})|^2.
\end{align}

% quadratic Zeeman effect
The quadratic Zeeman energy is induced by a linearly polarized microwave field as well as by an external magnetic field
as~\cite{Gerbier2006,Leslie2009}
\begin{align}
E_{\rm Z2} &= \int d{\bm r} \sum_{mm'}  \Psi_m^*({\bm r})\nonumber\\
&\times\left[q_B\left(\hat{\bm e}_B\cdot {\bm F}\right)^2+q_{\rm EM}\left(\hat{\bm e}_{\rm EM}\cdot {\bm F}\right)^2\right]_{mm'}
\Psi_{m'}({\bm r}),
\end{align}
where $\hat{\bm e}_{\rm EM}$ is the direction of the polarization of the microwave field,
$q_B=(g_F\mu_{\rm B}B)^2/E_{\rm hf}$ with $E_{\rm hf}$ being the hyperfine splitting energy,
and $q_{\rm EM}=-\hbar^2 \Omega^2/(4\delta)$ with $\Omega$ being the Rabi frequency and $\delta$ the detuning.
In this paper,
we take $\hat{\bm e}_{\rm EM}$ to be parallel to $\hat{\bm e}_B=\hat{z}$.
The quadratic Zeeman energy then becomes
\begin{align}
E_{\rm Z2} &= \int d{\bm r} \sum_{m}  qm^2 |\Psi_m({\bm r})|^2,
\label{eq:energy_q}
\end{align}
where $q=q_B+q_{\rm EM}$.

% dipole interaction (general)
The general form of the MDDI energy is given by
\begin{align}
E_{\rm dd} =& c_{\rm dd}\int d{\bm r} \int d{\bm r}' 
Q_{\nu\nu'}({\bm r}-{\bm r}') f_{\nu}({\bm r}) f_{\nu'}({\bm r}'),
\label{eq:energy_dd0}
\end{align}
where $c_{\rm dd}=\mu_0(g_F\mu_{\rm B})^2/(4\pi)$ with $\mu_0$ being the magnetic permeability of the vacuum,
and $Q_{\nu\nu'}({\bm r})$ is the dipole kernel
which will be given in Sec.~\ref{sec:dipole}.
Here and henceforth, the Greek subscripts that appear twice are to be summed over $x, y$, and $z$, 

% linear Zeeman effect
Note that if the atomic cloud is isolated in the vacuum,
the total magnetization along the external magnetic field is conserved,
as long as the dipolar relaxation (the spin relaxation due to the MDDI) can be ignored.
The dipolar relaxation is dominant for atoms with large magnetic dipole moments such as
$^{52}$Cr atoms~\cite{Hensler2003}, while it is negligible in BECs of alkali atoms~\cite{Chang2004}.
In the latter case, 
the linear Zeeman term can be eliminated if we choose the rotating frame of reference in spin space 
with the Larmor frequency $\omega_{\rm L}$
by transforming $\Psi_m({\bm r}, t)$ to $\Psi_m({\bm r},t)e^{-im\omega_{\rm L}t}$.
In the absence of the MDDI, the total energy functional~\eqref{eq:energy0} is invariant under this transformation
since $E_{\rm s}$ and $E_{\rm Z2}$ are invariant.
However, 
in the presence of the MDDI, $E_{\rm dd}$ is not invariant under this transformation,
and therefore we have to use the modified dipole kernel in the rotating frame, which will be given in Sec.~\ref{sec:dipole}.

\subsection{Nonlocal Gross-Pitaevskii equation}
% 3D GP equation
The mean-field dynamics of the system is governed by the nonlocal GPE:
\begin{align}
 i\hbar\frac{\partial}{\partial t} \Psi_{m}=& \frac{\delta (E-\mu N)}{\delta \Psi_m^*} \nonumber\\
 =& \left[-\frac{\hbar^2}{2M}\nabla^2 +U_{\rm trap}({\bm r}) -\mu + qm^2 + c_0 n \right]\Psi_{m}\nonumber\\
  & + \sum_{m'=0,\pm1}(c_1 f_\nu + c_{\rm dd}b_\nu)   (F_\nu)_{mm'}\Psi_{m'}
\label{eq:3dGP},
\end{align}
where $\mu$ is the chemical potential and 
\begin{align}
 b_\nu({\bm r}) = \int d{\bm r}' Q_{\nu\nu'}({\bm r}-{\bm r}')f_{\nu'}({\bm r}')
\end{align}
is the dipole field.

% 2D GP equation
We consider a harmonic trap 
\begin{align}
U_{\rm trap}({\bm r}) = M\left(\omega_1^2x_1^2 + \omega_2^2 x_2^2 + \omega_3^2 x_3^2\right)/2,
\end{align}
where $x_i\equiv\hat{\bm e}_i\cdot{\bm r}$ with $\hat{\bm e}_i$ being the trap axis.
Assuming $\omega_3 \gg \omega_{1,2}$, 
the wave function in the $\hat{\bm e}_3$ direction is approximated by a Gaussian
\begin{align}
h(x_3)=\frac{1}{(2\pi d^2)^{1/4}}\exp\left(-\frac{x_3^2}{4d^2}\right),
\end{align}
and the order parameter can be written as
\begin{align}
\Psi_m({\bm r},t)=\psi_m(x_1,x_2,t)h(x_3).
\end{align}
Multiplying Eq.~\eqref{eq:3dGP} by  $h(x_3)$ and integrating over $x_3$, we obtain the quasi-2D GPE:
\begin{align}
 &i\hbar\frac{\partial}{\partial t} \psi_{m}\nonumber\\
 &= \left[-\frac{\hbar^2}{2M}\nabla_{\perp}^2 + U_{\rm trap}^{\rm (2D)}({\bm r}_\perp) -\mu + qm^2 + \bar{c}_0 \bar{n} \right]\psi_{m}\nonumber\\
   &\ \ \ \ + \sum_{m'=0,\pm1}(\bar{c}_1 \bar{f}_\nu + \bar{c}_{\rm dd}\bar{b}_\nu)   (F_\nu)_{mm'}\psi_{m'}
\label{eq:2dGP},
\end{align}
where ${\bm r}_\perp = (x_1,x_2)$, $\nabla_\perp^2 = \frac{\partial^2}{\partial x_1^2} + \frac{\partial^2}{\partial x_2^2}$,
$U_{\rm trap}^{\rm (2D)}({\bm r}_\perp) = \frac{M}{2}\left(\omega_1^2x_1^2+\omega_2^2x_2^2\right)$,
$\bar{c}_{0,1} = c_{0,1}/\sqrt{4\pi d^2}$ and $\bar{c}_{\rm dd}=c_{\rm dd}/\sqrt{4\pi d^2}$.
We define the 2D number density, spin density, and dipole field by
\begin{align}
\bar{n} &= \int dx_3 n({\bm r}),\\
\bar{\bm f}&=\int dx_3 {\bm f}({\bm r}),\\
\bar{b}_\nu &= \int d^2 r'_\perp Q_{\nu\nu'}^{\rm (2D)}({\bm r}_\perp-{\bm r}'_\perp)\bar{f}_{\nu'}({\bm r}'_\perp),
\end{align}
respectively, where
\begin{align}
 &Q^{\rm (2D)}_{\nu\nu'}({\bm r}-{\bm r}') \nonumber\\
&= \sqrt{4\pi d^2} \iint dx_3 dx_3' h^2(x_3)h^2(x_3') Q_{\nu\nu'}({\bm r}-{\bm r}').
\label{eq:Q_3D_to_2D}
\end{align}
Using the Fourier transform of the dipole kernels
\begin{align}
Q_{\nu\nu'}({\bm r}) &= \sum_{\bm k}\tilde{Q}_{{\bm k}\nu\nu'} e^{i{\bm k}\cdot{\bm r}},\\
Q^{\rm (2D)}_{\nu\nu'}({\bm r}_\perp) &= \sum_{{\bm k}_\perp}\tilde{Q}^{\rm (2D)}_{{\bm k}_\perp\nu\nu'} e^{i{\bm k}_\perp\cdot{\bm r}_\perp},
\end{align}
with ${\bm k}_\perp = (k_1, k_2) \equiv ((\hat{\bm e}_1\cdot {\bm k}), (\hat{\bm e}_2\cdot {\bm k}))$, 
Eq.~\eqref{eq:Q_3D_to_2D} is rewritten as
\begin{align}
\tilde{Q}^{\rm (2D)}_{{\bm k}_\perp\nu\nu'} = \frac{d}{\sqrt{\pi}} \int dk_3 e^{-d^2 k_3^2} \tilde{Q}_{{\bm k}\nu\nu'},
\label{eq:Q_3D_to_2D_k}
\end{align}
where $k_3\equiv\hat{\bm e}_3\cdot {\bm k}$.

\subsection{Dipole kernel}
\label{sec:dipole}
Since the MDDI is long-ranged and couples spin and orbital degrees of freedom,
its kernel depends on the geometry of the condensate and on the frame of reference in spin space.
In this subsection, we discuss the dipole kernel in 3D and quasi-2D condensates 
both in the laboratory frame and the rotating frame at the Larmor frequency.

The dipole kernel in the laboratory frame of reference is given by
\begin{align}
 Q_{\nu\nu'}^{\rm (lab)}({\bm r}) = \frac{\delta_{\nu\nu'}-3\hat{r}_\nu\hat{r}_{\nu'}}{r^3},
\label{eq:Q_lab}
\end{align}
where $r=|{\bm r}|$, $\hat{r}={\bm r}/r$, and
its Fourier transform is given by
\begin{align}
 \tilde{Q}^{\rm (lab)}_{{\bm k}\nu\nu'}
&= -\frac{4\pi}{3}\left(\delta_{\nu\nu'}-3\hat{k}_{\nu}\hat{k}_{\nu'}\right),
\label{eq:Q_lab_k}
\end{align}
where $\hat{\bm k}={\bm k}/|{\bm k}|$.
To calculate the 2D dipole kernel in the laboratory frame, we expand $k_\nu$ as
$k_\nu = \sum_{i=1}^3 k_i(\hat{\bm e}_i)_\nu$ and substitute
Eq.~\eqref{eq:Q_lab_k} in Eq.~\eqref{eq:Q_3D_to_2D_k}.
Using the following integrals:
\begin{align}
&\frac{d}{\sqrt{\pi}}\int dk_3 e^{-d^2k_3^2}\frac{k_ik_j}{k_1^2+k_2^2+k_3^2} \nonumber\\
&=\left\{\begin{array}{ll}
G(k_\perp d)(\hat{\bm k}_\perp)_i(\hat{\bm k}_\perp)_j & (i,j=1,2),\\
1-G(k_\perp d) &(i=j=3),\\
0 &({\rm otherwise}),
\end{array}
\right.
\end{align}
with
$k_\perp = |{\bm k}_\perp|$, $\hat{\bm k}_\perp = {\bm k}_\perp/k_\perp$, $(\hat{\bm k}_\perp)_i = \hat{\bm k}_\perp \cdot \hat{\bm e}_i$, and
\begin{align}
G(k) \equiv 2 k e^{k^2} \int_k^\infty e^{-t^2}dt,
\end{align}
we obtain 
\begin{align}
 \tilde{Q}^{\rm (2D, lab)}_{{\bm k}_\perp\nu\nu'}
  &=  -\frac{4\pi}{3}\left[\delta_{\nu\nu'} -3 (\hat{\bm e}_3)_\nu(\hat{\bm e}_3)_{\nu'} \right]\nonumber \\
  &+ 4\pi G(k_\perp d)\left[(\hat{\bm k}_\perp)_\nu(\hat{\bm k}_\perp)_{\nu'} -(\hat{\bm e}_3)_\nu(\hat{\bm e}_3)_{\nu'}
\right],
\label{eq:Q_lab_2D_k}
\end{align}
where $(\hat{\bm k}_\perp)_\nu=\sum_{i=1,2}(\hat{\bm k}_\perp)_i(\hat{\bm e}_i)_\nu$.
It can be shown that $G(k)$ is a monotonically increasing function
that satisfies $G(0)=0$ and $G(\infty)=1$.

Next, we consider the dipole kernel in the rotating frame of reference.
Since the spin density vector in the rotating frame ${\bm f}^{\rm (rot)}$ is related to
that in the laboratory frame ${\bm f}^{\rm (lab)}$ as
\begin{align}
f_\nu^{\rm (lab)}= R_{\nu\nu'} f_{\nu'}^{\rm (rot)},
\label{eq:rotatingframe}
\end{align}
where
\begin{align}
R=
\begin{pmatrix}
 \cos\omega_{\rm L} t & -\sin\omega_{\rm L} t & 0 \\
 \sin\omega_{\rm L} t &  \cos\omega_{\rm L} t & 0 \\
0 & 0 &  1
\end{pmatrix},
\end{align}
the dipole kernel in the rotating frame of reference is given by $R^{\rm T}_{\nu\rho} Q_{\rho\rho'}^{\rm (lab)}({\bm r}) R_{\rho' \nu'}$,
where ${\rm T}$ denotes the matrix transpose.
When the Larmor precession is much faster than the dynamics caused by the MDDI,
we can use the time-averaged dipole kernel over the period of the Larmor precession~\cite{Kawaguchi2007}:
\begin{align}
Q_{\nu\nu'}^{\rm (rot)}({\bm r}) &= \left\langle R^{\rm T}_{\nu\rho} Q_{\rho\rho'}^{\rm (lab)}({\bm r}) R_{\rho' \nu'}\right \rangle\nonumber\\
&= -\frac{1}{2}\,\frac{1-3\hat{r}_z^2}{r^3}\left(\delta_{\nu\nu'}-3\delta_{z\nu}\delta_{z\nu'}\right),
\label{eq:Q_rot}
\end{align}
where $\langle\cdots\rangle$ represents the time average over $2\pi/\omega_{\rm L}$.
This is the case with the Berkeley experiment~\cite{Vengalattore2008},
where the MDDI energy is $c_{\rm dd}n/h\sim 1$~Hz and the Larmor frequency is $\omega_{\rm L}/(2\pi)=115~$kHz
under the magnetic field of $B=165$~mG.
This approximation ignores the spin-orbit coupling terms
that induce spin relaxation via the Einstein-de Haas effect~\cite{Kawaguchi2006a, Santos2006, Gawryluk2007},
and therefore the spin and orbital parts are decoupled in Eq.~\eqref{eq:Q_rot}.
The experimental finding
that the longitudinal magnetization is conserved over a period of 1~s in $^{87}$Rb BECs~\cite{Chang2004}
is consistent with this approximation.

The Fourier transform of $Q_{\nu\nu'}^{\rm (rot)}({\bm r})$ is given by
\begin{align}
 \tilde{Q}^{\rm (rot)}_{{\bm k}\nu\nu'} = \frac{2\pi}{3} \left(1-3\hat{k}_z^2\right)\left(\delta_{\nu\nu'}-3\delta_{z\nu}\delta_{z\nu'}\right).
 \label{eq:Q_rot_k}
\end{align}
Substituting Eq.~\eqref{eq:Q_rot_k} in Eq.~\eqref{eq:Q_3D_to_2D_k}, we obtain the time-averaged 2D dipole kernel in the rotating frame as
\begin{align}
 \tilde{Q}^{\rm (2D, rot)}_{{\bm k}_\perp\nu\nu'}
  &=  \left(\delta_{\nu\nu'}-3\delta_{z\nu}\delta_{z\nu'}\right)\tilde{\mathcal{Q}}_{{\bm k}_\perp},
\label{eq:Q_rot_2D_k}
\end{align}
where
\begin{align}
  \tilde{\mathcal{Q}}_{{\bm k}_\perp}=\frac{2\pi}{3}\left\{1-3(\hat{\bm e}_3)_z^2 -3G(k_\perp d)\left[(\hat{\bm k}_\perp)_z^2 - (\hat{\bm e}_3)_z^2 \right]\right\}.
\end{align}

\subsection{Ground-state phase diagram in the absence of MDDI}
\begin{figure}[t]
\includegraphics[width=0.9\linewidth]{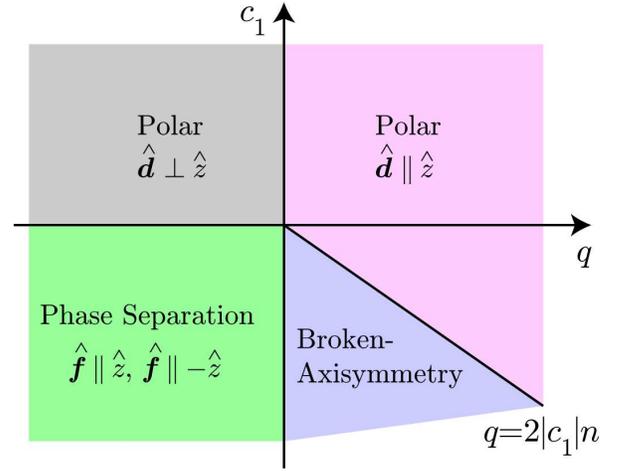}
\caption{(Color online) Phase diagram of a spin-1 BEC in the absence of the MDDI,
where the total longitudinal magnetization is fixed to be zero.
}
\label{fig:phase_diagram}
\end{figure}

%Ferromagnetic, Polar, and Broken-Axisymmetry Phases}
We briefly review ground-state properties of the spin-1 BEC
in a uniform system ($U_{\rm trap}=0$) in the absence of the MDDI.
The phase diagram is shown in Fig.~\ref{fig:phase_diagram}.
The ground-state spin configuration for the case of $q=0$ is uniform due to the kinetic term in Eq.~\eqref{eq:energy0};
for $c_1>0$, the condensate is unmagnetized (${\bm f}={\bm 0}$), while
for $c_1<0$, the condensate is fully magnetized ($|{\bm f}|=n$);
the former phase is called polar or antiferromagnetic, while the latter is called ferromagnetic~\cite{Ohmi1998, Ho1998}.

% ferromagnetic vs quadratic Zeeman
Though the direction of the spontaneous magnetization in the ferromagnetic phase is arbitrary for $q=0$,
the quadratic Zeeman effect
restricts the direction of magnetization~\cite{Stenger1998, Murata2007}.
Here we consider the case in which the total longitudinal magnetization is fixed to be zero.
Substituting $|\Psi_1| = |\Psi_{-1}| = \sqrt{(n-|\Psi_0|^2)/2}$ and minimizing Eq.~\eqref{eq:energy0},
one finds that the phase transition occurs at $q=2|c_1|n$.
For $q>2|c_1|n$, the quadratic Zeeman energy dominates the system and all atoms are condensed in the $m=0$ state,
while the ground state for $0<q<2|c_1|n$ is partially magnetized in the direction perpendicular to the external field.
The order parameter for the latter case is given by
\begin{align}
\begin{pmatrix}\Psi_1 \\ \Psi_0 \\ \Psi_{-1} \end{pmatrix}
=\frac{\sqrt{n} e^{i\phi}}{2}\begin{pmatrix}
  e^{-i\alpha}\sqrt{1-\frac{q}{2|c_1|n}} \\ \sqrt{2\left(1+\frac{q}{2|c_1|n}\right)} \\ e^{i\alpha}\sqrt{1-\frac{q}{2|c_1|n}}
\end{pmatrix},
\label{eq:OP_sp}
\end{align}
where $\phi$ and $\alpha$ are arbitrary real numbers.
The magnetization for this state is given by
\begin{align}
f_z&=0,\\
f_+&\equiv f_x+if_y = ne^{i\alpha}\sqrt{1-\left(\frac{q}{2|c_1|n}\right)^2}.
\end{align}
Hence, $\alpha$ denotes the direction of the magnetization in the $x$--$y$ plane,
and the amplitude of the magnetization depends on $q$.
Since the spontaneous magnetization breaks the axisymmetry around the external field,
this phase is called the broken-axisymmetry (BA) phase~\cite{Murata2007}.
The dynamics of this quantum phase transition has been investigated in Refs.~\cite{Saito2007a,Lamacraft2007,Uhlmann2007,Saito2007b,Sau2009,Swislocki2010}.
When $q<0$, the spin-polarized state with $m=1$ or $-1$
can minimize both the ferromagnetic interaction and the quadratic Zeeman energy.
To satisfy the conservation of the total longitudinal magnetization,
the phase separation of two domains with $f_z=1$ and $-1$ must occur.

The order parameter for the polar phase can be characterized with a unit vector $\hat{\bm d}$ as
\begin{align}
\sum_{m'}({\bm F}\cdot\hat{\bm d})_{mm'} \Psi_{m'} = 0.
\end{align}
For example, the order parameter for $\hat{\bm d}=\hat{z}$ is given by $(0,1,0)^{\rm T}$.
In the absence of the quadratic Zeeman effect, the direction of $\hat{\bm d}$ is arbitrary.
However, when $q$ is nonzero, the quadratic Zeeman effect restricts the direction of $\hat{\bm d}$:
$\hat{\bm d}=\hat{z}$ for $q>0$, while $\hat{\bm d}\perp \hat{z}$ for $q<0$.
The order parameter for the latter case is given by
\begin{align}
 \frac{e^{i\phi}}{\sqrt{2}}\begin{pmatrix} -\hat{d}_x+i\hat{d}_y \\ 0 \\ \hat{d}_x+i\hat{d}_y \end{pmatrix},
\end{align}
where $\phi$ is an arbitrary real number.

In the presence of the MDDI, the regions of the ferromagnetic phase and the BA phase expand 
since the MDDI favors ferromagnetic structure by developing spin textures.
The phase diagram in a quasi-2D system with a uniform spin structure is investigated in Ref.~\cite{Kjall2009},
where the phase boundary is predicted to depend on the geometry of the system and spin textures,
as will also be discussed below.

In the following sections, we consider a BEC of spin-1 $^{87}$Rb atoms
which is ferromagnetic with the scattering lengths of $a_0=101.8 a_{\rm B}$ and $a_2=100.4 a_{\rm B}$
where $a_{\rm B}$ is the Bohr radius~\cite{Kempen2002}.
The hyperfine g-factor of the spin-1 $^{87}$Rb atom is $g_F=-1/2$.
We choose $q$ to be in the region of the BA phase, i.e., $0<q<2|c_1|n$,
as in the Berkeley experiment~\cite{Vengalattore2008}.

\section{Dynamical Instability in a Uniform Quasi-2D System}
\label{sec:linearstability}
The linear stability of this system has previously been discussed in Refs.~\cite{Cherng2008, Lamacraft2008, Cherng2009}:
both the initial helix configuration and the MDDI induce the dynamical instability.
Here we investigate the stability of a spin helix with the MDDI in a uniform quasi-2D system ($U_{\rm trap}^{\rm (2D)}=0$).
For the case of a uniform spin structure, our result agrees with that in Ref.~\cite{Cherng2009}.
In this section, we consider only the quasi-2D system and omit the subscript $\perp$.
We assume that the external magnetic field is much stronger than the dipole field and use the dipole kernel~\eqref{eq:Q_rot_2D_k}.

Since $\alpha$ in Eq.~\eqref{eq:OP_sp} specifies the direction of the transverse magnetization,
the helical spin structure can be described as
\begin{align}
{\bm \psi} ({\bm r})
&= \sqrt{\bar{n}}
\begin{pmatrix}e^{-i{\bm \kappa}\cdot{\bm r}} & 0 & 0 \\ 0 & 1 & 0 \\ 0 & 0 & e^{i{\bm \kappa}\cdot{\bm r}}\end{pmatrix}
\begin{pmatrix} \zeta_1({\bm r}) \\ \zeta_0({\bm r})\\ \zeta_{-1}({\bm r}) \end{pmatrix}\nonumber \\
&\equiv\sqrt{\bar{n}} K {\bm \zeta}({\bm r}),
\label{eq:OP_helix_transformation}
\end{align}
where ${\bm \psi}=(\psi_1,\psi_0,\psi_{-1})^{\rm T}$, ${\bm \zeta}$ is a three-component spinor satisfying $\sum_{m=0,\pm1}|\zeta_m|^2=1$,
and $\bar{n} = \sqrt{2\pi d^2}n(x_1,x_2,x_3=0)$ is assumed to be constant.
Substituting Eq.~\eqref{eq:OP_helix_transformation} into quasi-2D GPE~\eqref{eq:2dGP}, we obtain
\begin{align}
 i\hbar \frac{\partial}{\partial t} {\bm \zeta}
 =& \left[ -\frac{\hbar^2}{2M}\left(\nabla - i \bm\kappa F_z\right)^2 -\mu
 + q F_z^2 +  \bar{c}_0\bar{n}\right.\nonumber \\
 &\ \ \  + \bar{c}_1\bar{n} ({\bm \zeta}^\dagger  F_\nu {\bm \zeta}) F_\nu 
 + \bar{c}_{\rm dd} \bar{n} \sum_{\bm k}e^{i{\bm k}\cdot{\bm r}}\mathcal{A}_{\bm k}\bigg]{\bm \zeta},
\label{eq:2dGP_helix}
\end{align}
where
\begin{align}
\mathcal{A}_{\bm k} \equiv &\int d{\bm r} e^{-i{\bm k}\cdot {\bm r}}
 \left[ -2\tilde{\mathcal{Q}}_{\bm k}({\bm \zeta}^\dagger  F_z {\bm \zeta})  F_z \right.\nonumber\\
 &+\frac{1}{2}\tilde{\mathcal{Q}}_{{\bm k}+{\bm\kappa}} ({\bm \zeta}^\dagger  F_+ {\bm \zeta}) F_-
 + \frac{1}{2}\tilde{\mathcal{Q}}_{{\bm k}-{\bm\kappa}} ({\bm \zeta}^\dagger  F_- {\bm \zeta}) F_+ 
\bigg],
\end{align}
with $F_+=F_-^\dagger=F_x+iF_y$, 
and we used the following relations:
\begin{align}
 K^\dagger F_\pm K &= e^{\pm i{\bm \kappa}\cdot{\bm r}}F_\pm, \\
 K^\dagger F_zK &= F_z,\\
 {\bm F}^2 \equiv F_\nu F_\nu &= \frac{F_+F_-+F_-F_+}{2}+F_z^2.
\end{align}

We consider a spin-helix state with $f_z=0$ (i.e., $|\zeta_1|=|\zeta_{-1}|$) as in the case of the Berkeley experiment~\cite{Vengalattore2008}.
Assuming that ${\bm \zeta}$ is uniform and satisfies $|\zeta_1|=|\zeta_{-1}|$,
we can rewrite Eq.~\eqref{eq:2dGP_helix} as
\begin{align}
 &\left[  
 \left(q+\frac{\hbar^2|{\bm \kappa}|^2}{2M}\right) F_z^2 +  \bar{c}_0\bar{n} \right.\nonumber \\
 &\ \ \  + (\bar{c}_1 + \bar{c}_{\rm dd}\tilde{\mathcal{Q}}_{\bm \kappa})\bar{n} 
 \sum_{\nu=x,y}({\bm \zeta}^\dagger F_\nu{\bm \zeta})F_\nu
\bigg]{\bm \zeta}
 = \mu{\bm \zeta},
\label{eq:2dGP_helix2}
\end{align}
where we used ${\bm \zeta}^\dagger F_z{\bm \zeta}=0$ and $\tilde{\mathcal{Q}}_{\bm \kappa}=\tilde{\mathcal{Q}}_{-\bm \kappa}$.
Note that if we replace $q$ with $q+\hbar^2|{\bm \kappa}|^2/(2M)$ and $\bar{c}_1$ with $\bar{c}_1 + \bar{c}_{\rm dd}\tilde{\mathcal{Q}}_{\bm \kappa}$.
Eq.~\eqref{eq:2dGP_helix2} takes the same form as the GPE~\eqref{eq:2dGP} for the stationary state [i.e., $\partial\psi_m/(\partial t)=0$]
in the absence of the MDDI and spin helix.
Therefore, as in the case of Eq.~\eqref{eq:OP_sp}, the solution for Eq.~\eqref{eq:2dGP_helix2} is given by
\begin{align}
{\bm \zeta}_0
= \frac{e^{i\phi}}{2}
\begin{pmatrix}
e^{-i\alpha}\sqrt{1-\tilde{q}_{\bm \kappa}}\\
\sqrt{2(1+\tilde{q}_{\bm \kappa})} \\
e^{i\alpha}\sqrt{1-\tilde{q}_{\bm \kappa}}
\end{pmatrix},
\label{eq:stationary}
\end{align}
where
\begin{align}
\tilde{q}_{\bm \kappa} = \frac{q+\hbar^2 |{\bm \kappa}|^2/(2M)}{2 \bar{n}(|\bar{c}_1|-\bar{c}_{\rm dd}\tilde{\mathcal{Q}}_{\bm\kappa})}.
\label{eq:def_tildeq}
\end{align}
The chemical potential and the energy per particle are given by
\begin{align}
\mu&= \frac{q}{2} + \frac{\hbar^2 |{\bm \kappa}|^2}{4M} + \bar{n}(\bar{c}_0+\bar{c}_1+\bar{c}_{\rm dd} \tilde{\mathcal{Q}}_{\bm\kappa}),\\
 E_{\bm \kappa} &= \frac{q}{2} + \frac{\hbar^2 |{\bm \kappa}|^2}{4M} + \frac{1}{2}\bar{n}(\bar{c}_0+\bar{c}_1+\bar{c}_{\rm dd} \tilde{\mathcal{Q}}_{\bm\kappa}).
\end{align}

Note that when the magnetic field is applied parallel to the 2D plane ($\hat{\bm e}_3=\hat{y}$),
the helical spin structure can have lower energy than the uniform spin structure.
On the other hand, when the magnetic field is perpendicular to the 2D plane ($\hat{\bm e}_3=\hat{z}$),
the uniform spin structure has the lowest energy.
With $d=1.0~\mu$m, $n=2.3\times 10^{14}~{\rm cm}^{-3}$, and $\hat{\bm e}_3=\hat{y}$,
$E_{\bm \kappa}$ takes a minimum of $E_{{\bm \kappa}_0}-E_{\bm 0}=-0.067~{\rm Hz}\times h$
at ${\bm \kappa}_0\simeq 2\pi/(138~\mu{\rm m}) \hat{z}$.

The Bogoliubov equation for a spinor BEC is derived by substituting
\begin{align}
{\bm\zeta} = {\bm \zeta}_0 
+ \sum_{\bm k} [{\bm u}_{\bm k}e^{i({\bm k}\cdot{\bm r}-\epsilon_{\bm k}t/\hbar)} + {\bm v}_{\bm k}^* e^{-i({\bm k}\cdot{\bm r}-\epsilon_{\bm k}^*t/\hbar)}]
\end{align}
into Eq.~\eqref{eq:2dGP_helix} and linearizing the result with respect to ${\bm u}_{\bm k}$ and ${\bm v}_{\bm k}$.
Here ${\bm u}_{\bm k}$ and ${\bm v}_{\bm k}$ are three component spinors,
and therefore, we have the $6\times 6$ eigenvalue matrix equation:
\begin{align}
 \begin{pmatrix} M_{\bm k} & N_{\bm k} \\ -N^*_{-\bm k} & -M^*_{-\bm k} \end{pmatrix}
\begin{pmatrix} {\bm u}_{\bm k} \\ {\bm v}_{\bm k} \end{pmatrix}
=
\epsilon_{\bm k}\begin{pmatrix} {\bm u}_{\bm k} \\ {\bm v}_{\bm k} \end{pmatrix},
\label{eq:6by6_BdG}
\end{align}
where $M_{\bm k}$ and $N_{\bm k}$ are $3\times 3$ matrices defined by
\begin{align}
M_{\bm k} &= \frac{\hbar^2}{2M}\left({\bm k}-{\bm \kappa}F_z\right)^2 -\mu + qF_z^2
 + \bar{c}_0 \bar{n} \left[1 + ({\bm\zeta}_0 {\bm \zeta}_0^\dagger )\right]\nonumber\\
 &+ \bar{c}_1 \bar{n} \left[({\bm \zeta}_0^\dagger  F_\nu{\bm \zeta}_0)F_\nu  + F_\nu({\bm \zeta}_0{\bm \zeta}_0^\dagger ) F_\nu\right]\nonumber\\
 & -2 \bar{c}_{\rm dd} \bar{n}\left[
  \tilde{\mathcal{Q}}_{\bm 0} ({\bm \zeta}_0^\dagger  F_z {\bm \zeta}_0)  F_z
  + \tilde{\mathcal{Q}}_{\bm k} F_z({\bm \zeta}_0 {\bm \zeta}_0^\dagger ) F_z \right]\nonumber\\
 & + \frac{\bar{c}_{\rm dd} \bar{n}}{2}
 \left[\tilde{\mathcal{Q}}_{\bm\kappa} ({\bm \zeta}_0^\dagger  F_+ {\bm \zeta}_0) F_-
 + \tilde{\mathcal{Q}}_{\bm\kappa} ({\bm \zeta}_0^\dagger  F_- {\bm \zeta}_0) F_+ \right.\nonumber\\
 &\left.\ \ \ \ \ + \tilde{\mathcal{Q}}_{\bm k+\bm\kappa} F_-({\bm \zeta}_0{\bm \zeta}_0^\dagger ) F_+ 
 + \tilde{\mathcal{Q}}_{\bm k-\bm\kappa} F_+({\bm \zeta}_0{\bm \zeta}_0^\dagger ) F_-\right],\\
 N_{\bm k} &= 
 \bar{c}_0 \bar{n} ({\bm\zeta}_0 {\bm \zeta}_0^{\rm T})
 + \bar{c}_1 \bar{n} F_\nu({\bm \zeta}_0{\bm \zeta}_0^{\rm T} ) F^{\rm T}_\nu \nonumber\\
 &-2 \bar{c}_{\rm dd} \bar{n}\tilde{\mathcal{Q}}_{\bm k} F_z({\bm \zeta}_0 {\bm \zeta}_0^{\rm T} ) F_z \nonumber\\
 & + \frac{\bar{c}_{\rm dd} \bar{n}}{2}
 \left[ \tilde{\mathcal{Q}}_{\bm k+\bm\kappa} F_-({\bm \zeta}_0{\bm \zeta}_0^{\rm T} ) F_- 
 + \tilde{\mathcal{Q}}_{\bm k-\bm\kappa} F_+({\bm \zeta}_0{\bm \zeta}_0^{\rm T} ) F_+ \right].
\end{align}
Here,
$({\bm \zeta}{\bm \zeta}^{\dagger})$ and
$({\bm \zeta}{\bm \zeta}^{\rm T})$ are $3\times 3$ matrices
whose $(m,m')$ components are given by $\zeta_m \zeta_{m'}^*$ and $\zeta_m \zeta_{m'}$, respectively,
while ${\bm \zeta}^\dagger F_\nu {\bm \zeta}=\sum_{mm'} \zeta_m^* (F_\nu)_{mm'}\zeta_{m'}$ is a scalar.

Since the $6\times 6$ matrix in Eq.~\eqref{eq:6by6_BdG} is not Hermitian (though $M_{\bm k}$ and $N_{\bm k}$ are Hermitian), the eigenvalue can be complex.
When the eigenvalue has a nonzero imaginary part,
the corresponding mode becomes dynamically unstable and exponentially grows or decays.
When $\epsilon_{\bm k}$ is a real eigenvalue with wavenumber ${\bm k}$, $-\epsilon_{\bm k}$ is also an eigenvalue of Eq.~\eqref{eq:6by6_BdG} with wavenumber $-{\bm k}$;
the sign of $\epsilon_{\bm k}$ is determined
so that the corresponding eigenstate satisfies ${\bm u}_{\bm k}^\dagger {\bm u}_{\bm k} - {\bm v}_{\bm k}^\dagger {\bm v}_{\bm k}=1$.
In the presence of the energy dissipation, 
the mode for $\epsilon_{\bm k}<0$ is energetically unstable due to the Landau instability.

When ${\bm \kappa}={\bm 0}$, one of the Bogoliubov modes can be obtained analytically.
By rewriting ${\bm u}$ and ${\bm v}$ as
\begin{align}
 \begin{pmatrix} u_1\\ u_0 \\ u_{-1} \end{pmatrix} 
&= \frac{1}{\sqrt{2}}\begin{pmatrix} -1 & i & 0 \\ 0 & 0 & \sqrt{2} \\ 1 & i & 0 \end{pmatrix} 
 \begin{pmatrix} u_x\\ u_y \\ u_z \end{pmatrix}, \\
 \begin{pmatrix} v_1\\ v_0 \\ v_{-1} \end{pmatrix} 
&= \frac{1}{\sqrt{2}}\begin{pmatrix} -1 & i & 0 \\ 0 & 0 & \sqrt{2} \\ 1 & i & 0 \end{pmatrix} 
 \begin{pmatrix} v_x\\ v_y \\ v_z \end{pmatrix}, 
\end{align} 
the $(u_x, v_x)$ mode is decoupled from the other two modes, and the eigenvalue equation \eqref{eq:6by6_BdG} is reduced to the $2\times 2$ matrix equation given by
\begin{align}
 \begin{pmatrix} f_{\bm k} & g_{\bm k} \\ -g_{\bm k} & -f_{\bm k} \end{pmatrix} \begin{pmatrix} u_{x{\bm k}} \\ v_{x{\bm k}} \end{pmatrix}
 = \epsilon_{\bm k} \begin{pmatrix} u_{x{\bm k}} \\ v_{x{\bm k}} \end{pmatrix},
\end{align}
where
\begin{align}
f_{\bm k} &=  \frac{\hbar^2 k^2}{2M} + \frac{q}{2} 
- \bar{c}_{\rm dd} \bar{n} \tilde{\mathcal{Q}}_{\bm 0} + \bar{c}_{\rm dd} \bar{n} \frac{-1+3\tilde{q}_{\bm 0}}{2} \tilde{\mathcal{Q}}_{\bm k}, \\
g_{\bm k} &=  |\bar{c}_1|\bar{n}\tilde{q}_{\bm 0} + \bar{c}_{\rm dd} \bar{n} \frac{-3+\tilde{q}_{\bm 0}}{2} \tilde{\mathcal{Q}}_{\bm k}.
\end{align}
The eigenvalue of this spin mode is obtained as
\begin{align}
 \epsilon_{\bm k}^2 =& (f_{\bm k} + g_{\bm k}) (f_{\bm k} - g_{\bm k}) \nonumber \\
 =&\left[\frac{\hbar^2 k^2}{2M} + q - \bar{c}_{\rm dd} \bar{n} (1-\tilde{q}_{\bm 0})(2\tilde{\mathcal{Q}}_{\bm k} +\tilde{\mathcal{Q}}_{\bm 0} ) \right]\nonumber\\
  &\times\left[\frac{\hbar^2 k^2}{2M} + \bar{c}_{\rm dd} \bar{n}(1+\tilde{q}_{\bm 0})(\tilde{\mathcal{Q}}_{\bm k}-\tilde{\mathcal{Q}}_{\bm 0}) \right],
\label{eq:eigenfreq}
\end{align}
which agrees with the result obtained in Ref.~\cite{Cherng2009}.

\begin{figure}[t]
\includegraphics[width=\linewidth]{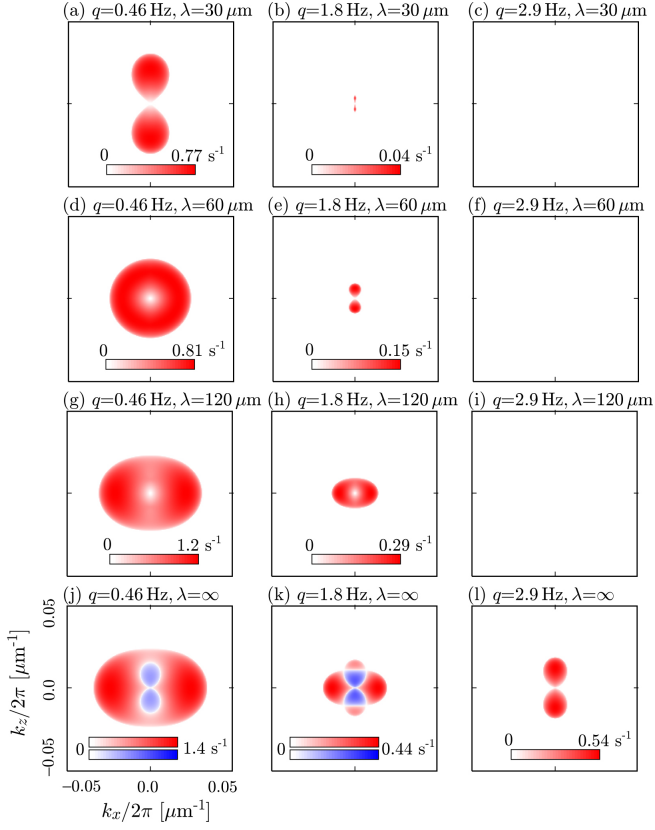}
\caption{(Color) Quadratic Zeeman energy $q$ and helical pitch $\lambda$ dependence of the dynamical and Landau instabilities.
The abscissa and ordinate show $k_x/2\pi$ and $k_z/2\pi$, respectively.
Shown are $|{\rm Im}\,\epsilon_{\bm k}|$ (red) and $-{\rm Re}\,\epsilon_{\bm k}$ (blue) which correspond to the dynamical and Landau instabilities, respectively;
they are calculated for $(\hat{\bm e}_1, \hat{\bm e}_2, \hat{\bm e}_3)=(\hat{z},\hat{x},\hat{y})$, ${\bm \kappa}=(2\pi/\lambda)\hat{z}$, 
$n(x_3=0)=2.3\times 10^{14}~{\rm cm}^{-3}$ and $d=1.0~\mu$m.
No instability is found for (c), (f), and (i).
}
\label{fig:BdG_ene}
\end{figure}

For ${\bm \kappa}\neq{\bm 0}$, we numerically solve the Bogoliubov equation in Eq.~\eqref{eq:6by6_BdG}.
The results for the case of $\hat{\bm e}_3 = \hat{y}$ and ${\bm \kappa}=(2\pi/\lambda)\hat{z}$
are summarized in Fig.~\ref{fig:BdG_ene},
where we show the distributions of $|{\rm Im}\,\epsilon_{\bm k}/h|$ (red) and $-{\rm Re}\,\epsilon_{\bm k}/h$ (blue) in the momentum space
which correspond to the dynamical instability and Landau instability, respectively.
Since the eigenvalue equation~\eqref{eq:6by6_BdG} is the $6\times 6$ matrix equation, there are three independent solutions.
We have numerically confirmed that
the lowest-energy mode becomes unstable in some parameter regime (see Fig.~\ref{fig:BdG_ene}),
and the other two modes are stable.
The eigenvalue of the lowest-energy mode continuously approaches Eq.~\eqref{eq:eigenfreq} as $\lambda\to\infty$.

\begin{figure}[t]
\includegraphics[width=\linewidth]{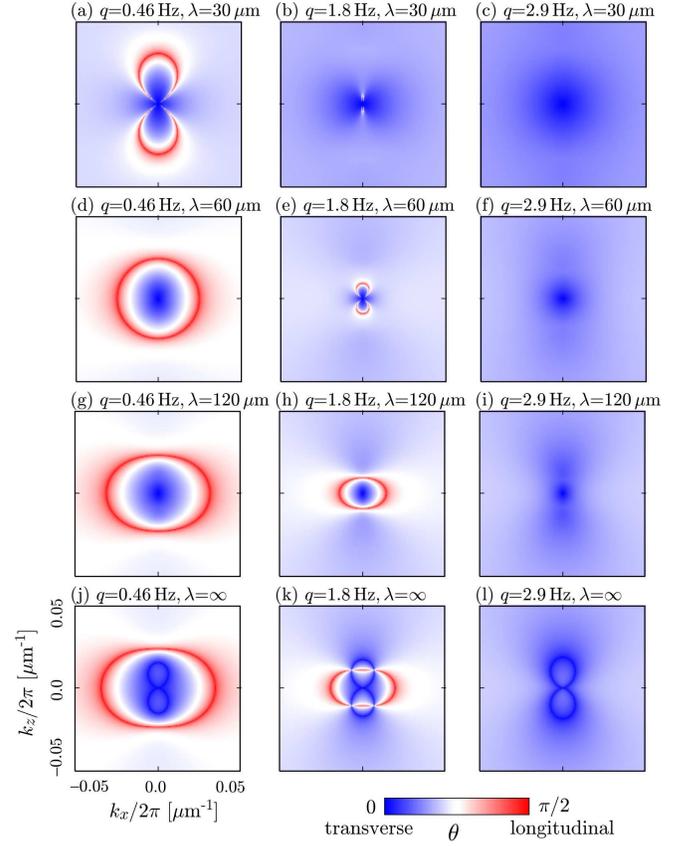}
\caption{(Color) Direction $\theta$ of the magnetic fluctuation defined in Eq.~\eqref{eq:def_theta} for the lowest-energy modes.
Each panel corresponds to that in Fig.~\ref{fig:BdG_ene}.
The parameters are the same as those in Fig.~\ref{fig:BdG_ene}.
}
\label{fig:BdG_mode}
\end{figure}

The magnetic fluctuations are analyzed as follows.
The eigenmode
${\bm w} = {\bm u}_{\bm k}e^{i({\bm k}\cdot{\bm r}-\epsilon_{\bm k}t/\hbar)} + {\bm v}_{\bm k}^* e^{-i({\bm k}\cdot{\bm r}-\epsilon_{\bm k}^*t/\hbar)}$
induces fluctuations in the transverse and longitudinal magnetizations as
\begin{align}
\Delta f_y &\equiv ({\bm \zeta}_0+{\bm w})^\dagger F_y({\bm \zeta}_0+{\bm w}) \nonumber \\
 &\simeq -\sqrt{1+\tilde{q}_{\bm \kappa}}\,{\rm Im}(w_1-w_{-1}),
\label{eq:delta_fy}\\
\Delta f_z &\equiv ({\bm \zeta}_0+{\bm w})^\dagger F_z({\bm \zeta}_0+{\bm w}) \nonumber \\
 &\simeq  \sqrt{1-\tilde{q}_{\bm \kappa}}\,{\rm Re}(w_1-w_{-1}),
\label{eq:delta_fz}
\end{align}
which can be rewritten in the form of
\begin{align}
\Delta f_{y,z}= A_{y,z} \sin[{\bm k}\cdot{\bm r}-({\rm Re}\,\epsilon_{\bm k})t/\hbar + \chi_{y,z}].
\end{align}
We may use this result to define the direction of the magnetic fluctuation $\theta$ as
\begin{align}
 \theta \equiv \arctan\frac{A_z}{A_y}.
\label{eq:def_theta}
\end{align}
We will call the magnetic fluctuation transverse if $0 \le \theta< \pi/4$ and longitudinal if $\pi/4< \theta \le \pi/2$.
For an eigenmode with a real eigenvalue,
${\bm u}$ and ${\bm v}$ are also real, and $\theta$ is calculated as
\begin{align}
\tan \theta = \sqrt{\frac{1-\tilde{q}_{\bm \kappa}}{1+\tilde{q}_{\bm \kappa}}}\left|\frac{u_1+v_1-u_{-1}-v_{-1}}{u_1-v_1-u_{-1}+v_{-1}}\right|.
\end{align}
On the other hand, for an eigenmode with a complex eigenvalue,
we rewrite the eigenmode as
${\bm u}={\bm u}'+i{\bm u}''$ and ${\bm v}={\bm v}'+i{\bm v}''$, where ${\bm u}', {\bm u}'', {\bm v}'$ and ${\bm v}''$ are real,
and we obtain
\begin{align}
&\tan^2\theta =\frac{1-\tilde{q}_{\bm \kappa}}{1+\tilde{q}_{\bm \kappa}}\nonumber\\
&\times
\frac{(u'_{1}+v'_{1}-u'_{-1}-v'_{-1})^2+(u''_{1}+v''_{1}-u''_{-1}-v''_{-1})^2}{(u'_{1}-v'_{1}-u'_{-1}+v'_{-1})^2+(u''_{1}-v''_{1}-u''_{-1}+v''_{-1})^2}.
\label{eq:theta_for_complex_mode}
\end{align}
Equation~\eqref{eq:theta_for_complex_mode}
does not depend on the overall phase of the eigenmode, i.e.,
it is invariant under $({\bm u},{\bm v}) \to e^{i\chi}({\bm u},{\bm v})$ for an arbitrary real $\chi$.

For the case of ${\bm \kappa}={\bm 0}$, $\theta$ for the $(u_x, v_x)$ mode is analytically obtained both for real and imaginary eigenvalues as
\begin{align}
\tan^2\theta &= \frac{1-\tilde{q}_{\bm 0}}{1+\tilde{q}_{\bm 0}}\left|\frac{f_{\bm k}-g_{\bm k}}{f_{\bm k}+g_{\bm k}}\right|\nonumber\\
&=\frac{1-\tilde{q}_{\bm 0}}{1+\tilde{q}_{\bm 0}}
\left|
\frac{\frac{\hbar^2 k^2}{2M} + \bar{c}_{\rm dd} \bar{n}(1+\tilde{q}_{\bm 0})(\tilde{\mathcal{Q}}_{\bm k}-\tilde{\mathcal{Q}}_{\bm 0})}
{\frac{\hbar^2 k^2}{2M} + q - \bar{c}_{\rm dd} \bar{n} (1-\tilde{q}_{\bm 0})(2\tilde{\mathcal{Q}}_{\bm k} +\tilde{\mathcal{Q}}_{\bm 0} )}
\right|.
\end{align}
We also calculate $\theta$ numerically for ${\bm \kappa}\neq{\bm 0}$.
In Fig.~\ref{fig:BdG_mode},
we plot $\theta$ for the lowest-energy modes
which become unstable in the red regions of Fig.~\ref{fig:BdG_ene}.

The Bogoliubov analysis shown in Figs.~\ref{fig:BdG_ene} and \ref{fig:BdG_mode}
suggests that for large $q \ (\gtrsim E_{\rm dd})$ the MDDI favors a helical spin structure,
since the dynamical instability that exists at $\lambda=\infty$ is suppressed as $\lambda$ decreases, as shown in Fig.~\ref{fig:BdG_ene}.
Actually, the ground-state spin structure at large $q$ is a spin helix as discussed in Sec.~\ref{sec:GS}.
On the other hand, for small $q \ (\lesssim E_{\rm dd})$,  fluctuations in both longitudinal and transverse magnetizations become dynamically unstable.
The Landau instability at $\lambda=\infty$
reflects the fact that $E_{\bm \kappa}$ has the minimum for $\kappa_z \neq 0$:
$\theta$ corresponding to the Landau instability is small [Figs.~\ref{fig:BdG_mode} (j) and (k)],
implying that the instability is caused by the fluctuations of transverse magnetizations.

Although the Bogoliubov analysis indicates that the MDDI favors nonuniform spin structures,
it is insufficient to account for the Berkeley experiment~\cite{Vengalattore2008}:
as discussed in Ref.~\cite{Cherng2009},
the minimum wave length of the unstable modes ($\sim 30~\mu$m) is about three times larger than 
that of the observed magnetic pattern ($\lambda_{\rm exp}\sim 10~\mu$m).

\section{Spin Dynamics in a Trapped System}
\label{sec:sim_pancake}
To take into account the effects of the trapping potential, initial conditions, noises, and nonlinearity,
we perform numerical simulations of the full 3D GPE with the dipole kernel in Eq.~\eqref{eq:Q_rot}.
We first consider a large axisymmetric pancake-shaped BEC to eliminate the effect of anisotropy of the trap,
and then discuss the case of the Berkeley experiment~\cite{Vengalattore2008} in the next section.

%% GP equation, initial state --
We consider a BEC of $N=1.0\times 10^7$ atoms in a harmonic trap with frequencies $(\nu_x, \nu_y, \nu_z) = (4.2, 420, 4.2)$~Hz.
The corresponding peak density is $n({\bm 0})=2.0\times 10^{14}~{\rm cm}^{-3}$ and the Thomas-Fermi (TF) radii are $(r_x, r_y, r_z)=(140, 1.4, 140)~\mu{\rm m}$.
The width along the $y$ direction is smaller than
the spin healing length $\xi_{\rm sp}\equiv\hbar/\sqrt{2M|c_1|n({\bm 0})}=2.8~\mu{\rm m}$
and the dipole healing length $\xi_{\rm dd}\equiv \hbar/\sqrt{2M c_{\rm dd} n({\bm 0})} = 9.5~\mu{\rm m}$.
The magnetic field $B=165$~mG is applied in the $z$ direction.

In the simulation,
we first calculate a stationary state, $\Psi^{\rm (ini)}({\bm r})$, polarized in the $m=-1$ state by using the imaginary-time propagation method,
i.e., we have solved the stationary state of Eq.~\eqref{eq:3dGP} by replacing $t$ with $-it$.
Then the initial state is given as
\begin{align}
 \begin{pmatrix} \Psi_1 \\ \Psi_0 \\ \Psi_{-1} \end{pmatrix}
= \mathcal{N} |\Psi^{\rm (ini)}({\bm r})| \frac{e^{i\gamma}}{2}
\begin{pmatrix}
e^{-i({\bm \kappa}\cdot{\bm r}+\alpha)}\frac{1+\beta}{2}  \sqrt{1-\tilde{q}_{\bm \kappa}}\\
\sqrt{2(1+\tilde{q}_{\bm \kappa})} \\
e^{i({\bm \kappa}\cdot{\bm r}+\alpha)}\frac{1-\beta}{2} \sqrt{1-\tilde{q}_{\bm \kappa}}
\end{pmatrix},
\label{eq:OP_pancake_initial}
\end{align}
where ${\bm \kappa}=\kappa \hat{z}$, 
$\tilde{q}_{\bm \kappa}$ is defined in Eq.~\eqref{eq:def_tildeq} whose
denominator is replaced by $2n({\bm 0})(|c_1|-c_{\rm dd}\tilde{\mathcal{Q}}_{\bm \kappa})$,
and $\mathcal{N}$ is a normalization constant;
the parameters $\alpha$, $\beta$ and $\gamma$ are introduced to simulate
fluctuations in the transverse magnetization,
the longitudinal magnetization, and the overall phase, respectively,
due to the quantum and thermal fluctuations as well as experimental noises.
They are assumed to take on real random numbers independently on each grid
and to obey the Gaussian distribution with variance $\sigma_{\alpha,\beta,\gamma}$.
In the following calculation, we choose $\sigma_\alpha=\sigma_\gamma=0.1$ and $\sigma_\beta=0.03$.
The initial noise dependence of the spin dynamics is discussed in Sec.~\ref{sec:discussion}.

\subsection{Effect of the nonuniform density}
\begin{figure}[th]
\includegraphics[width=\linewidth]{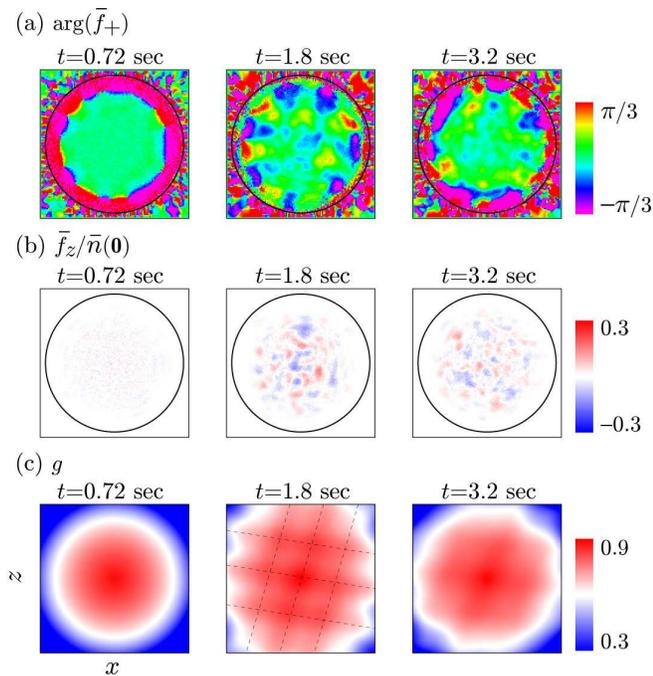}
\caption{
(Color) Time evolution of (a) the transverse magnetization, (b) the longitudinal magnetization, and (c) the spin correlation function~\eqref{eq:def_correlation}
in the absence of the MDDI.
The initial spin configuration is uniform and $q/h=2.9$~Hz.
The solid circles in (a) and (b) represent the TF radius ($140~\mu$m) at $y=0$.
The size of each panel is $310~\mu{\rm m}\times 310~\mu{\rm m}$.
The peak-to-peak distance of the correlation function at $t=1.8$~s is about $70~\mu$m.
}
\label{fig:pancake_cdd0}
\end{figure}

We first consider the spin dynamics without the MDDI.
Interestingly, a periodic pattern of the transverse magnetization
develops even in the case of $c_{\rm dd}=0$ and $\kappa=0$.
Figure~\ref{fig:pancake_cdd0} shows the time evolution of the magnetic structure for $q/h=2.9$~Hz.
Shown are (a) the direction of the transverse magnetization ${\rm arg}(\bar{f}_+)$, 
(b) longitudinal magnetization $\bar{f}_z/\bar{n}({\bm 0})$, and 
(c) spin correlation function
\begin{align}
g({\bm r}_\perp)\equiv \frac{\int d^2 r'_\perp \bar{f}_+({\bm r}'_\perp+{\bm r}_\perp)\bar{f}_-({\bm r}'_\perp)}{\int d^2 r'_\perp \bar{n}({\bm r}'_\perp+{\bm r}_\perp)\bar{n}({\bm r}'_\perp)},
\label{eq:def_correlation}
\end{align}
where $\bar{n}({\bm r}_\perp)=\int dy n(x,y,z)$ is the column density and
$\bar{\bm f}({\bm r}_\perp)=\int dy {\bm f}(x,y,z)$ is the column spin density.
As shown in Fig.~\ref{fig:pancake_cdd0}, 
the magnetic pattern develops from a uniform spin structure,
and the periodic pattern appears in the spin correlation function at $t=1.8$~s, which is destroyed in the further time evolution.
This instability is due to the nonuniform density profile in the trapping potential.

In the present case of a $^{87}$Rb BEC,
the density-density interaction dominates the system ($c_0 n \gg |c_1| n, q$),
and therefore the density distribution $n({\bm r})$ is determined independently of the spin configuration.
Using Eq.~\eqref{eq:OP_sp}, 
the ground-state order parameter in the local density approximation (LDA) is given by
\begin{align}
 \bm\Psi_{\rm LDA}({\bm r}) = 
 \frac{\sqrt{n({\bm r})}}{2}
 \begin{pmatrix} \sqrt{1-\frac{q}{2|c_1|n({\bm r})}} \\ \sqrt{2\left[1+\frac{q}{2|c_1|n({\bm r})}\right]} \\ \sqrt{1-\frac{q}{2|c_1|n({\bm r})}} \end{pmatrix}
 \label{eq:LDA1}
\end{align}
for $n({\bm r})>q/(2|c_1|)$, and
\begin{align}
 \bm\Psi_{\rm LDA}({\bm r}) = 
 \sqrt{n({\bm r})}
 \begin{pmatrix} 0 \\ 1 \\ 0\end{pmatrix}
\end{align}
for $n({\bm r})<q/(2|c_1|)$.
The transverse magnetization per particle for $\bm\Psi_{\rm LDA}$ distributes according to
\begin{align}
 \frac{|f_+({\bm r})|}{n({\bm r})} = \left\{\begin{array}{lll}
	  \sqrt{1-\left[\frac{q}{2|c_1|n({\bm r})}\right]^2}  &{\rm for} & n({\bm r})>\frac{q}{2|c_1|},\\
		 0 &{\rm for} & n({\bm r})<\frac{q}{2|c_1|},
\end{array}
 \right.
\end{align}
while that of the initial state \eqref{eq:OP_pancake_initial} is constant (except for the contribution from the noise term):
\begin{align}
 \frac{|f_+({\bm r})|}{n({\bm r})} 
 = \sqrt{1-\left[\frac{q}{2|c_1|n({\bm 0})}\right]^2}.
 \label{eq:LDA2}
\end{align}
Therefore, the initial sate \eqref{eq:OP_pancake_initial} is not stationary at the low density region,
and the instability grows from the periphery [Fig.~\ref{fig:pancake_cdd0} (a) 0.72~s].
In the course of time evolution, the amplitude of the local magnetization oscillates at the periphery
and then the fluctuations begin to penetrate into the central region at $t\sim1.8$~s.
These fluctuations induce a periodic pattern in the correlation function as shown in the snapshot at $t=1.8$~s in Fig.~\ref{fig:pancake_cdd0} (c),
which lasts for about 1~s and eventually dissolves [Fig.~\ref{fig:pancake_cdd0} (c) 3.2~s].
The peak-to-peak distance in the correlation function at $t=1.8$~s is about $70~\mu$m.

\subsection{Effect of spin current}
\begin{figure}[th]
\includegraphics[width=\linewidth]{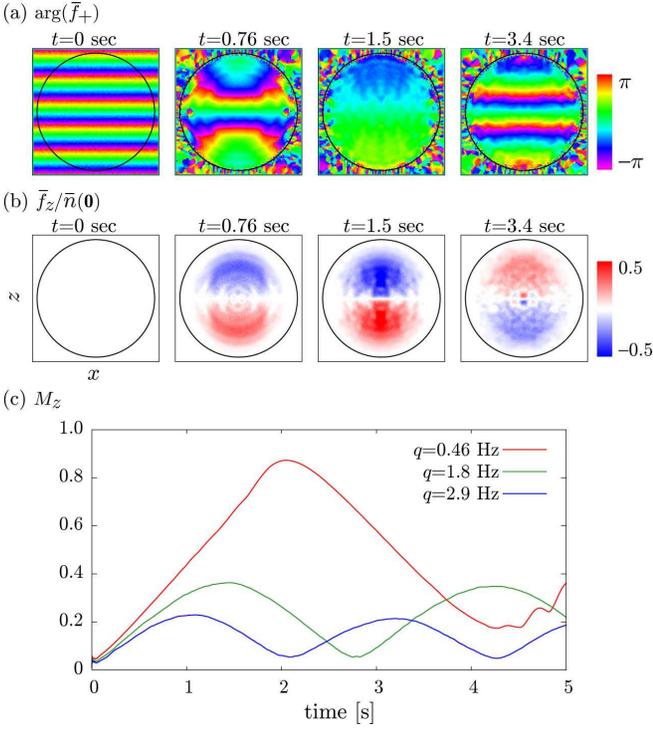}
\caption{
(Color) Spin dynamics
in the absence of the MDDI starting with a spin helix of $\lambda=60~\mu{\rm m}$.
Shown are the snapshots of (a) the transverse magnetization and (b) the longitudinal magnetization for $q/h=1.8$~Hz,
and (c) the time evolution of $M_z$ for $q/h=0.46$~Hz (red), 1.8~Hz (green), and 2.9~Hz (blue).
The size of each panel in (a) and (b) is $310~\mu{\rm m}\times 310~\mu{\rm m}$.
}
\label{fig:pancake_cdd0_helix}
\end{figure}

The effect of the trapping potential becomes more prominent when the initial condition is a spin helix.
Figure~\ref{fig:pancake_cdd0_helix} shows the spin dynamics in the absence of the MDDI for $q/h=1.8$~Hz
starting with a spin helix with $\lambda\equiv 2\pi/\kappa=60~\mu$m,
where Figs.~\ref{fig:pancake_cdd0_helix} (a) and (b) are the snapshots of 
the transverse and longitudinal magnetizations, respectively,
and Fig.~\ref{fig:pancake_cdd0_helix} (c) shows
the time evolution of the amplitude of the longitudinal magnetization $M_z\equiv \int d{\bm r} |f_z({\bm r})| / N$.
Note here that the helical spin structure induces a spin current of the longitudinal magnetization defined by
\begin{align}
{\bm j}^{\rm spin}_z &=\frac{\hbar}{2Mi}\sum_{m,m'}(F_z)_{mm'}[\Psi_m^*\nabla\Psi_{m'} - (\nabla\Psi_m^*)\Psi_{m'}].
\label{eq:spincurrent}
\end{align}
The initial spin helix induces the spin current of ${\bm j}^{\rm spin}_z(t=0)=-\hbar \kappa n \hat{z}/(2M)$,
and hence,
the longitudinal magnetization is accumulated at the top and bottom of the condensate as shown in Figs.~\ref{fig:pancake_cdd0_helix} (b) and (c).
Moreover, the spin current is reflected at the edge of the condensate, generating an interference pattern
as shown in the snapshots at $t=0.76$~s in Figs.~\ref{fig:pancake_cdd0_helix} (a) and (b).
In the time evolution, the spin helix unwinds [Fig.~\ref{fig:pancake_cdd0_helix} (a) 1.5 s]
and winds again in the opposite sense [Fig.~\ref{fig:pancake_cdd0_helix} (a) 3.4 s].
In this dynamics, the direction of the spin current is also inverted.
In the long-time scale of a few seconds, the spin texture oscillates between helix and anti-helix configurations, leading to 
the oscillations of $M_z$ as shown in Fig.~\ref{fig:pancake_cdd0_helix} (c).
The local longitudinal magnetization can become larger for the smaller quadratic Zeeman energy, and therefore, the period of the oscillations becomes longer.

\subsection{Effect of MDDI}
\begin{figure}[th]
\includegraphics[width=\linewidth]{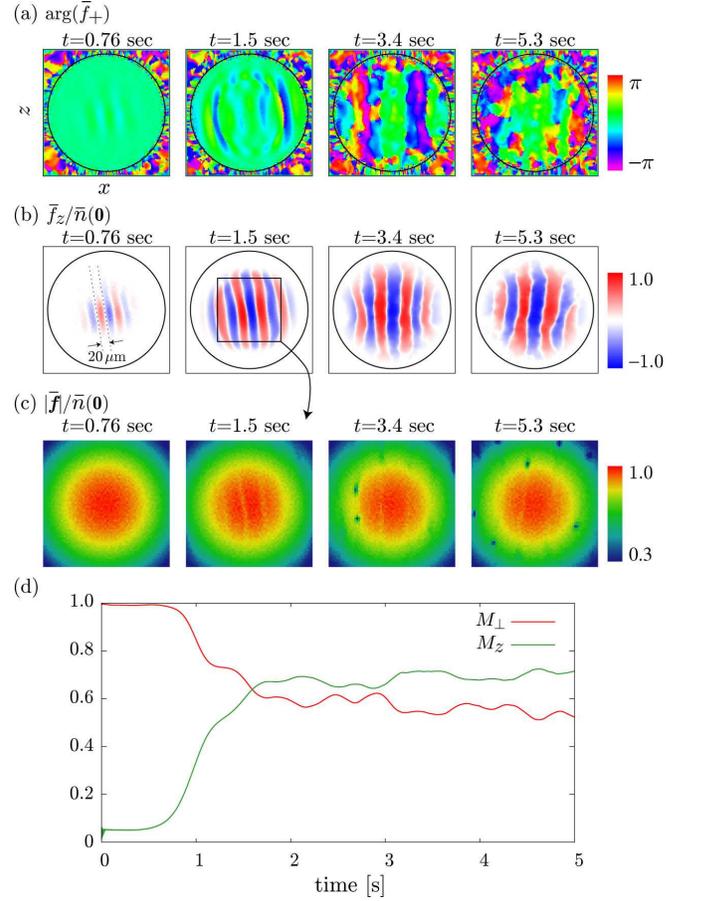}
\caption{
(Color) Spin dynamics
in the presence of the MDDI for $q/h=0.46$~Hz starting with a uniform spin structure.
The corresponding unstable mode is shown in Figs.~\ref{fig:BdG_ene} (j) and \ref{fig:BdG_mode} (j).
Shown are snapshots of (a) transverse magnetization, (b) longitudinal magnetization,
and (c) amplitude of the magnetization,
and (d) time developments of $M_\perp$ (red) and $M_z$ (green).
The size of each panel is $310~\mu{\rm m}\times 310~\mu{\rm m}$ in (a) and (b),
and $155~\mu{\rm m}\times 155~\mu{\rm m}$ in (c).
}
\label{fig:pancake_cdd1-1}
\end{figure}

Next, we consider the effect of the MDDI.
Figure~\ref{fig:pancake_cdd1-1} shows the result for $\kappa=0$ and $q/h=0.46$~Hz,
where Figs.~\ref{fig:pancake_cdd1-1} (a), (b), and (c) are the snapshots of transverse magnetization ${\rm arg}(\bar{f}_+)$, 
longitudinal magnetization $\bar{f}_z/\bar{n}({\bm 0})$, and amplitude of the magnetization $|\bar{\bm f}|/{\bar n}({\bm 0})$, respectively,
and Fig.~\ref{fig:pancake_cdd1-1} (d) shows the time evolution of the amplitude of the transverse magnetization per particle
$M_\perp\equiv \int d{\bm r}\sqrt{f_x^2+f_y^2}/N$ and that of the longitudinal magnetization $M_z$.
The Bogoliubov spectrum shown in Figs.~\ref{fig:BdG_ene} (j) and \ref{fig:BdG_mode} (j) predicts that the spin wave along the $x$ direction grows
with the most unstable wave length of $41~\mu$m in the time scale of $h/|{\rm Im}\,\epsilon_{\bm k}|\sim 0.74$~s.
The direction of the magnetic fluctuation for the most unstable mode is $\tan\theta=0.96$, i.e., both the transverse and longitudinal fluctuations grow.
The numerical result shown in Fig.~\ref{fig:pancake_cdd1-1} agrees well with the Bogoliubov analysis:
the stripe pattern of the longitudinal magnetization grows over the time scale of $\sim$ 1~s
where the domain size is $20~\mu$m.
The fluctuation in the transverse magnetization also grows.
As the magnetic domain of the longitudinal magnetization develops,
the magnetization $|\bar{\bm f}|$ decreases at the domain walls [Fig.~\ref{fig:pancake_cdd1-1} (c) 1.5~s].
In the course of time evolution, these domain wall are destroyed by generating pairs of polar-core vortices.
The polar-core vortex is a topologically stable spin vortex whose core is un-magnetized, i.e., filled with the polar state.
Each blue dot in the snapshots at $t=3.4$~s and 5.3~s in Fig.~\ref{fig:pancake_cdd1-1} (c) indicates
the core of a polar-core vortex.
The obtained stripe structure is relatively stable with the domain size gradually becoming larger 
as shown in the snapshot at $t=5.3$~s in Fig.~\ref{fig:pancake_cdd1-1} (b).
%The energy of the MDDI released by developing the staggered domains is transfered to the energy of the domain walls and their fluctuation.

\begin{figure}[th]
\includegraphics[width=\linewidth]{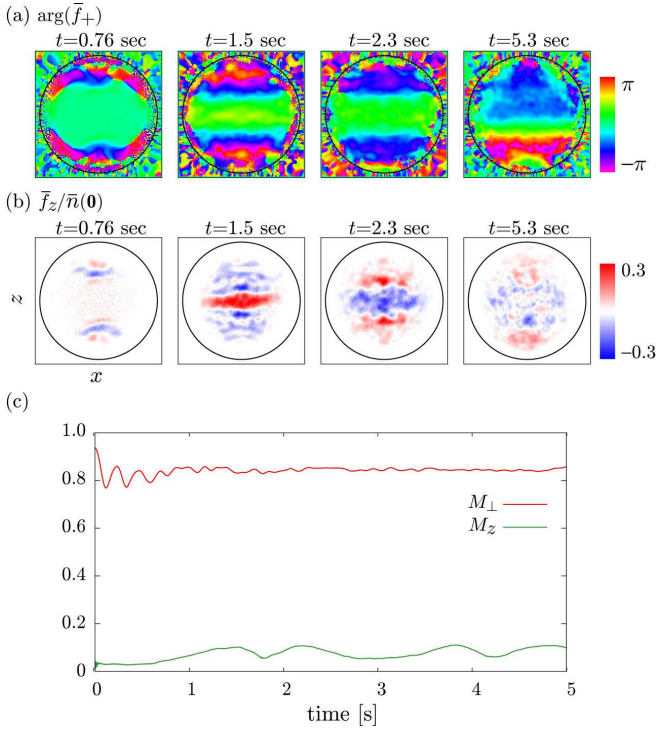}
\caption{
(Color) Spin dynamics
in the presence of the MDDI for $q/h=2.9$~Hz starting with a uniform spin structure.
The corresponding unstable mode is shown in Figs.~\ref{fig:BdG_ene} (l) and \ref{fig:BdG_mode} (l).
Shown are snapshots of (a) transverse magnetization and (b) longitudinal magnetization,
and (c) time evolution of $M_\perp$ (red) and $M_z$ (green).
The size of each panel in (a) and (b) is $310~\mu{\rm m}\times 310~\mu{\rm m}$.
}
\label{fig:pancake_cdd1-2}
\end{figure}

On the other hand, when $q/h=2.9$~Hz and $\kappa=0$,
the spin-wave mode of the transverse magnetization along the $z$ direction becomes unstable [Figs.~\ref{fig:BdG_ene} (l) and \ref{fig:BdG_mode} (l)].
Figure~\ref{fig:pancake_cdd1-2} shows the numerical results for $q/h=2.9$~Hz and $\kappa=0$, in agreement with the Bogoliubov analysis.
Due to the large quadratic Zeeman energy, the longitudinal magnetization cannot grow for $q/h=2.9$~Hz [Fig.~\ref{fig:pancake_cdd1-2} (c)].

\begin{figure}[th]
\includegraphics[width=\linewidth]{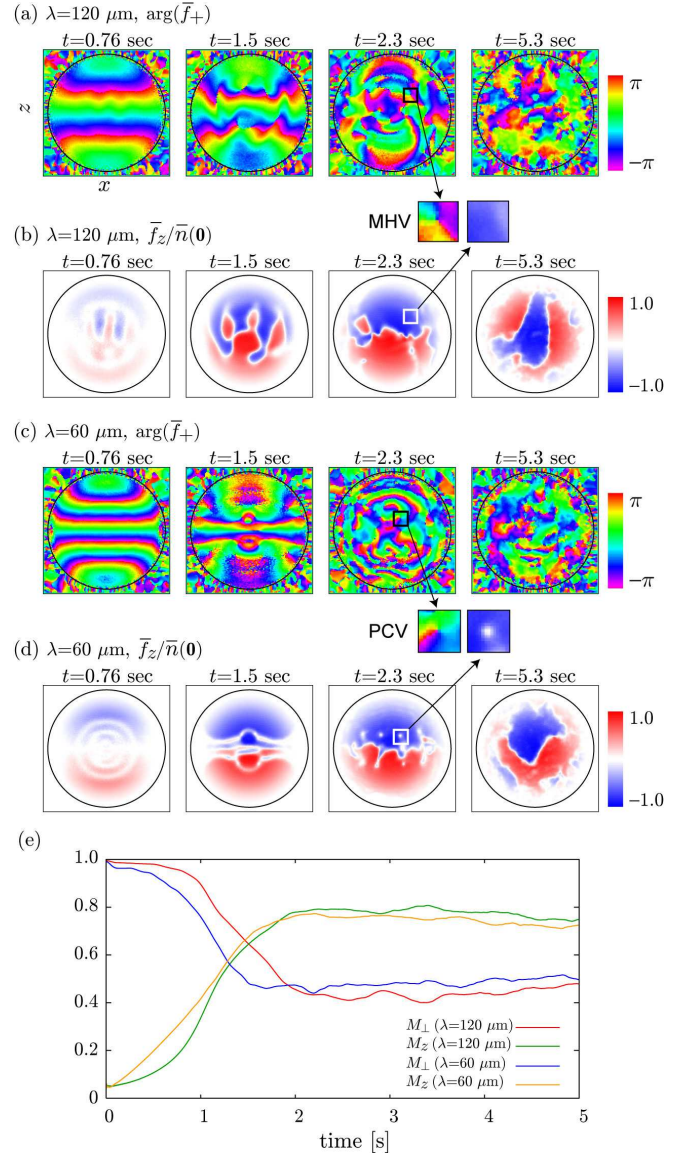}
\caption{
(Color) Spin dynamics
in the presence of the MDDI for $q/h=0.46$~Hz starting with a spin helix with a pitch of (a), (b) $\lambda=120~\mu$m and (c), (d) $\lambda=60~\mu$m.
Shown are snapshots of (a), (c) transverse magnetization and (b), (d) longitudinal magnetization.
(e) Time evolution of $M_\perp$ and $M_z$.
The size of each panel in (a)--(d) is $310~\mu{\rm m}\times 310~\mu{\rm m}$.
Examples for the MH vortex (MHV) and polar-core vortex (PCV) are enlarged at $t=2.3$~s in (a), (b) and (c), (d), respectively,
where the vortex core of MH vortex is magnetized and that of polar-core vortex is un-magnetized.
In both cases of MH and polar-core vortices, the transverse magnetization changes $2\pi$ around the vortex core.
}
\label{fig:pancake_cdd1_helix-1}
\end{figure}

The initial spin helix changes the distribution of the unstable mode in the Bogoliubov spectrum (Fig.~\ref{fig:BdG_ene}),
as well as induces the spin current of the longitudinal magnetization.
Figure~\ref{fig:pancake_cdd1_helix-1} shows the spin dynamics
starting with a spin helix with (a), (b) $\lambda=120~\mu$m and with (c), (d) $\lambda=60~\mu$m for $q/h=0.46$~Hz.
For $\lambda=120~\mu$m, the instability along the $x$ direction grows.
At the same time, the longitudinal magnetization is accumulated at the top and bottom of the condensate,
leading to the magnetic pattern shown in the snapshots at $t=1.5$~s in Figs.~\ref{fig:pancake_cdd1_helix-1} (a) and (b).
Then, helical structure is completely destroyed.
On the other hand, for $\lambda=60~\mu$m, 
the large initial spin current dominates the initial spin dynamics.
Different from the case of Fig.~\ref{fig:pancake_cdd0_helix},
the pitch of the helix becomes smaller and smaller, and finally the helical structure is destroyed
by generating pairs of polar-core vortices [Fig.~\ref{fig:pancake_cdd1_helix-1} (c) 1.5 s].
In both cases of $\lambda=120~\mu$m and $60~\mu$m, the local longitudinal magnetization substantially increases [Fig.~\ref{fig:pancake_cdd1_helix-1} (e)], and
the magnetic domains tend to elongate in the $z$ direction.
The polar-core vortices are located mainly at the domain wall of the longitudinal magnetization
for both $\lambda=60~\mu$m and $120~\mu$m.
We have also observed Mermin-Ho (MH) vortices for $\lambda=120~\mu$m,
where the core of the MH vortex is magnetized and the direction of the transverse magnetization changes $\pm 2\pi$ around it.
The examples of the MH and polar-core vortices are indicated in Fig.~\ref{fig:pancake_cdd1_helix-1}.

\begin{figure}[th]
\includegraphics[width=\linewidth]{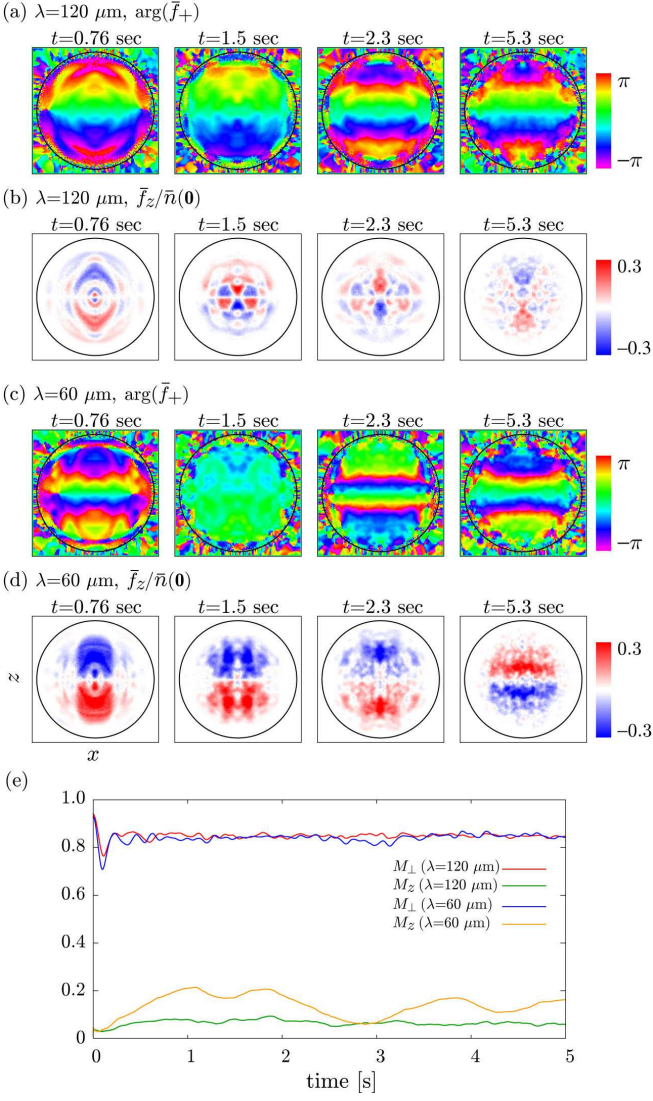}
\caption{
(Color) Spin dynamics
in the presence of the MDDI for $q/h=2.9$~Hz starting with a spin helix with a pitch of (a), (b) $\lambda=120~\mu$m and (c), (d) $\lambda=60~\mu$m.
Shown are snapshots of (a), (c) transverse magnetization and (b), (d) longitudinal magnetization.
(e) Time evolution of $M_\perp$ and $M_z$.
The size of each panel in (a)--(d) is $310~\mu{\rm m}\times 310~\mu{\rm m}$.
The sense of twisting of the spin helix dose not change in (a), while it is reversed in (c) at $t=2.3$~s.
}
\label{fig:pancake_cdd1_helix-2}
\end{figure}

Figure~\ref{fig:pancake_cdd1_helix-2} shows the spin dynamics for $q/h=2.9$~Hz
starting from a spin helix with a pitch of (a), (b) $\lambda=120~\mu$m and (c), (d) $\lambda=60~\mu$m.
Although the BEC for these parameters is stable in an infinite quasi-2D system [Fig.~\ref{fig:BdG_ene} (f) and (i)],
magnetic structures evolve due to the effect of the nonuniform density and the spin current.
In the case for a large helical pitch of $\lambda=120~\mu$m,
the helical structure of transverse magnetization is stable [Fig.~\ref{fig:pancake_cdd1_helix-2} (a)].
In this case, a regular magnetic pattern of the longitudinal magnetization emerges as an interference pattern of the spin current [Fig.~\ref{fig:pancake_cdd1_helix-2} (b)].
On the other hand, in the case of $\lambda=60~\mu$m,
the large initial spin current induces the untwisting and re-twisting of the helix as in the case of Fig.~\ref{fig:pancake_cdd0_helix},
although this dynamics is no longer periodic [Fig.~\ref{fig:pancake_cdd1_helix-2} (e)].

\section{Comparison with the Berkeley Experiment}
\label{sec:compare}

\subsection{Numerical results}
Now we consider the system of the Berkeley experiment~\cite{Vengalattore2008}.
The difference from the previous subsection is that
(i) the trapping potential is elliptical and elongated along the direction of the magnetic field,
(ii) the initial state is fully magnetized [$|{\bm f}({\bm r})|=n({\bm r})$]
while $|{\bm f}({\bm r})|=n({\bm r})\sqrt{1-\tilde{q}_{\bm k}^2}< n({\bm r})$ for the order parameter~\eqref{eq:OP_pancake_initial},
and (iii) the spin helix is evolved by applying a magnetic field gradient during a period of 5--8~ms.

In accordance with the Berkeley experiment, we consider a BEC of $N=2.3\times 10^6$ atoms in a harmonic trap with frequencies $(\nu_x,\nu_y,\nu_z)=(39, 440, 4.2)$~Hz.
The peak density is $n({\bm 0})=2.8\times 10^{14}~{\rm cm}^{-3}$ and the TF radii are $(r_x, r_y, r_z)=(18, 1.6, 169)~\mu$m.
The spin and dipole healing lengths are $\xi_{\rm sp}=2.3~\mu$m and $\xi_{\rm dd}=8.0~\mu$m, respectively.

For the initial state, we prepare a spin-polarized state in the $x$ direction as
\begin{align}
 \begin{pmatrix} \psi_1 \\ \psi_0 \\ \psi_{-1} \end{pmatrix}
 = \mathcal{N}|\Psi^{\rm (ini)}({\bm r})|e^{i\gamma} \left[\begin{pmatrix} e^{-i\alpha}\frac{1+\beta}{2} \\ \frac{1}{\sqrt{2}} \\ e^{i\alpha}\frac{1-\beta}{2} \end{pmatrix}
 +\delta\begin{pmatrix} \frac{1}{2}\\ -\frac{1}{\sqrt{2}}\\ \frac{1}{2}\end{pmatrix}
\right],
\label{eq:initial}
\end{align}
where $\Psi^{(\rm ini)}({\bm r}), \mathcal{N}, \alpha, \beta$, and $\gamma$ are the same as those appearing in Eq.~\eqref{eq:OP_pancake_initial}.
We introduce $\delta$ to simulate fluctuations in the amplitude of the magnetization.
Here we assume that $\delta$ takes on a complex random number independently on each grid
and obeys the Gaussian distribution with variance $\sigma_\delta$.
In the following numerical calculation, 
we choose $\sigma_\alpha=\sigma_\gamma=0.1$ and $\sigma_\beta=\gamma_\delta=0.03$.
We prepare a helical structure with pitch $\lambda$ in the real-time evolution
by applying a field gradient of $dB/dz=h/(|g_F|\mu_{\rm B}\lambda\tau_p)$ during a period of $\tau_p=5$~ms.

\begin{figure}[th]
\includegraphics[width=\linewidth]{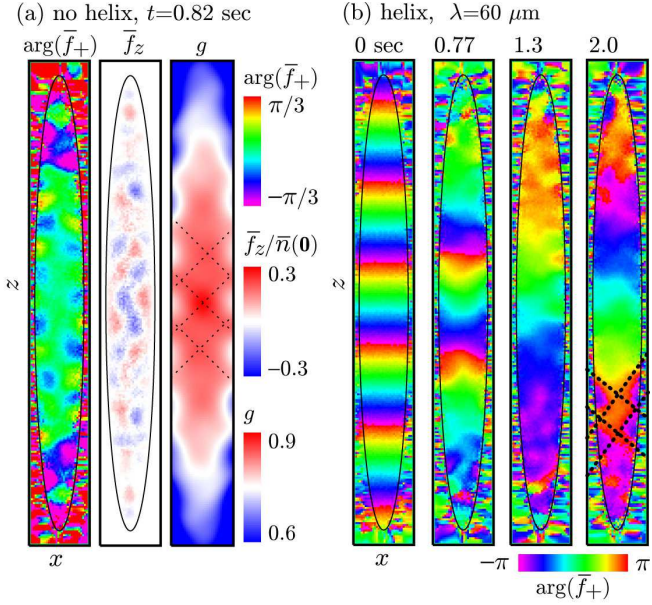}
\caption{
(Color) 
(a) Snapshots of spin configuration at $t=0.82$~s in the absence of the MDDI starting with a uniform spin configuration for $q/h=2.9$~Hz,
where each panel shows the distribution of transverse magnetization ${\rm arg}(\bar{f}_+)$ (left),
longitudinal magnetization $\bar{f}_z/\bar{n}_{\rm tot}({\bm 0})$ (middle), 
and spin correlation function $g$ (right).
The peak-to-peak distance of the correlation function is about $25~\mu$m.
(b) Time development of transverse magnetization in the absence of the MDDI starting with a spin helix of pitch $\lambda=60~\mu$m for $q/h=1.8$~Hz.
The domain size of the checkerboard pattern in (b) is about $20~\mu$m.
The solid curves in the distribution of longitudinal and transverse magnetizations represent the TF boundaries at $y=0$.
The size of each panel is $45~\mu{\rm m}\times 359~\mu{\rm m}$.
Note that the color scales of ${\rm arg}(\bar{f}_+)$ in (a) and (b) are different.
}
\label{fig:B_cdd0}
\end{figure}

We first consider the spin dynamics for $\kappa=0$ and $c_{\rm dd}=0$.
Since the TF radius in the $x$ direction is small,
the effect of the nonuniform density becomes more prominent than the previous case.
In addition, the deviation from the LDA stationary state discussed in Eqs.~\eqref{eq:LDA1}--\eqref{eq:LDA2} becomes larger
since we start with a fully-magnetized state.
Hence, the time and length scales of a checkerboard pattern become smaller;
the periodic pattern emerges spontaneously at $t\sim0.7$~s 
and the peak-to-peak distance of the correlation function is about $25~\mu$m.
The snapshots of the spin configuration at $t=0.82$~s are shown in Fig.~\ref{fig:B_cdd0} (a).
The periodic pattern eventually dissolves in a few seconds.

When we start from a spin helix with $c_{\rm dd}=0$, the longitudinal magnetization is accumulated at the top and bottom of the condensate
as in the case of a pancake-shaped trap.
However, different from the pancake-shaped trap,
the checkerboard pattern of the transverse magnetization
with the domain size $15 \sim 20~\mu$m appears at the top and bottom edges of the condensate for $t\ge 0.7$~s
due to the strong effects of the nonuniform density and the interference of the spin current.
The oscillations between the helix and anti-helix structures are observed as in the case of the pancake-shaped trap (Fig.~\ref{fig:pancake_cdd0_helix}),
and the checkerboard pattern disappears during these oscillations.

\begin{figure}[th]
\includegraphics[width=\linewidth]{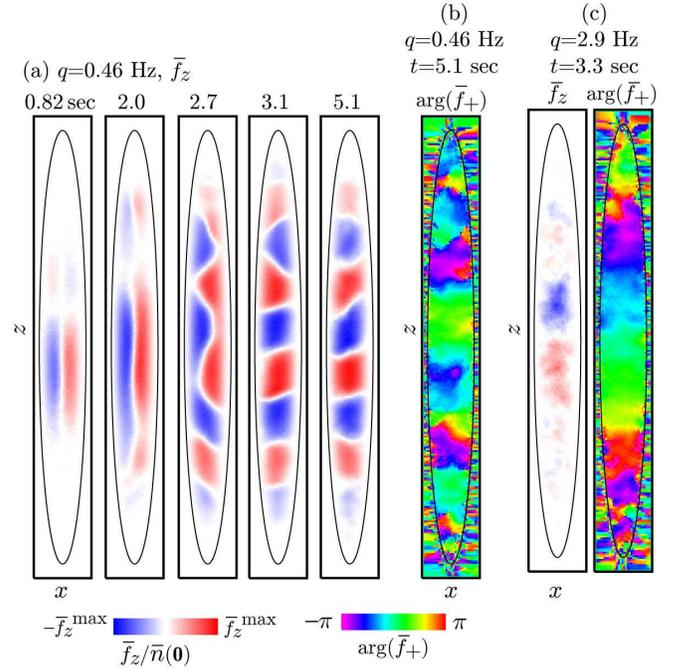}
\caption{
(Color) Magnetic structures developed from a uniform spin configuration in the presence of the MDDI at (a), (b) $q/h=0.46$~Hz and (c) 2.9~Hz,
where (a) and the left panel in (c) show the distributions of the longitudinal magnetization,
while (b) and the right panel in (c) show the direction of the transverse magnetization.
The longitudinal magnetization is scaled with $\bar{f}_z^{\rm max}=1.0$ in (a), and $\bar{f}_z^{\rm max}=0.6$ in the left panel of (c).
The transverse magnetization forms a helical structure in both cases for $q/h=0.46$ and 2.9~Hz.
The size of each panel is $45~\mu{\rm m}\times 359~\mu{\rm m}$.
}
\label{fig:B_cdd1}
\end{figure}

Figure~\ref{fig:B_cdd1} shows the spin dynamics in the presence of the MDDI for $\kappa=0$.
For $q/h=0.46$~Hz, 
fluctuations of the longitudinal magnetization grow in the $x$ direction [Fig.~\ref{fig:B_cdd1} (a) 0.82~s],
in agreement with the Bogoliubov analysis [Figs.~\ref{fig:BdG_ene} (j) and \ref{fig:BdG_mode} (j)].
However, in the further time evolution, the domains are aligned along the $z$ direction [Fig.~\ref{fig:B_cdd1} (a) 3.1~s],
because the long domain wall parallel to the $z$ direction is energetically unfavorable.
This domain structure seems to be a micro-canonical equilibrium state for small $q$,
and lasts for a long time [Fig.~\ref{fig:B_cdd1} (a) 5.1~s].
In the equilibrium state, the transverse magnetization forms a helix pattern along the $z$ direction [Figs.~\ref{fig:B_cdd1} (b)].
At large $q$, the checkerboard pattern first develops 
in the short time scale of $t\lesssim 1$~s as in the case of $c_{\rm dd}=0$.
Then, the effect of the MDDI on the spin dynamics becomes more prominent in the longer time scale.
For $q/h=2.9$~Hz, the fluctuations of magnetizations grow mainly in the transverse direction
and form a helix along the $z$ direction after a few seconds [Fig.~\ref{fig:B_cdd1} (c)].

\begin{figure}[th]
\includegraphics[width=0.9\linewidth]{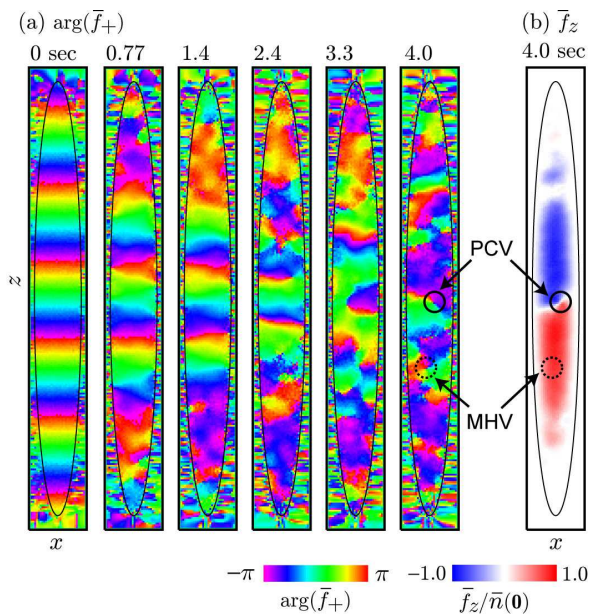}
\caption{
(Color) (a) Time development of the direction of the transverse magnetization with the MDDI starting with a spin helix of pitch $\lambda=60~\mu$m for $q/h=1.8$~Hz.
(b) Snapshot of the longitudinal magnetization at $t=4.0$~s.
The size of each panel is $45~\mu{\rm m}\times 359~\mu{\rm m}$.
The solid circles and dotted circles in (a) 4.0~s and (b) indicate the locations of the polar-core vortex (PCV) and MH vortex (MHV), respectively.
}
\label{fig:B_cdd1_helix}
\end{figure}

Finally, we show the result for the case of the Berkeley experiment~\cite{Vengalattore2008},
i.e., the spin dynamics in the presence of the MDDI starting with a spin helix.
Figure~\ref{fig:B_cdd1_helix} shows the result for $\lambda=60~\mu$m and $q/h=1.8$~Hz.
In the short time scale of $t\lesssim 1$~s,
the longitudinal magnetization is accumulated at the top and bottom of the condensate due to the spin current.
At the same time, the helical structure is modulated at the top and bottom of the condensate 
due to the effects of the nonuniform density and interference of the spin current.
The pitch of the helix becomes smaller and smaller, and the helix is eventually destroyed due to the MDDI.
In this dynamics, both polar-core vortices and MH vortices are generated spontaneously as indicated in Fig.~\ref{fig:B_cdd1_helix}.

\subsection{Comparison with the experiment}

Here we summarize the agreements and disagreements between the Berkeley experiment~\cite{Vengalattore2008}
and our numerical simulation.

In agreement with the experiment,
the uniform spin structure is stable over a few hundreds of milliseconds.
However, unlike the experimental results,
the spin helix is also stable within the time scale of a few hundreds of milliseconds
with and without the MDDI.
In the longer time scale of a few seconds, 
magnetic patterns develop from both the uniform spin structure and the spin helix
due to the trapping potential and the MDDI.
Even in the absence of the MDDI,
the nonuniform density distribution and interference of the spin current induce the checkerboard pattern (Fig.~\ref{fig:B_cdd0}).
However, the domain size observed in this dynamics is at least three times larger than that observed in the experiment,
and the checkerboard pattern eventually dissolves in our simulations.
The MDDI does not stabilize this periodic pattern.
The pattern induced by the MDDI is sensitive to the quadratic Zeeman energy.
When the MDDI dominates the quadratic Zeeman energy, the staggered domain of the longitudinal magnetization appears [Figs.~\ref{fig:B_cdd1} (a)].
On the other hand, when the quadratic Zeeman energy dominates the MDDI, magnetization is almost transverse
and forms a spin helix [Figs.~\ref{fig:B_cdd1} (c)].
The length scale of these structures is of the order of some tens of micrometers and much larger than that observed in the experiment.

Apart from the time and length scales,
there are several discrepancies in the property of the magnetic structure between the experiment and our calculation.
First, in our calculation, the instability is always accompanied by the emergence of the local longitudinal magnetization,
whereas in the experiment the longitudinal magnetization is much smaller than the transverse one.
In particular, when we start with a spin helix,
the longitudinal magnetization grows rapidly and becomes comparable to the transverse magnetization.
Second, the spin structure induced by the MDDI is sensitive to the external magnetic field,
whereas the spin dynamics observed in the Berkeley experiment is insensitive to the quadratic Zeeman energy for 0.8 $<q/h<$ 4~Hz.
Third, the growth rate of the checkerboard pattern (due to the nonuniform density and the spin current in our case) is insensitive to the pitch of the initial helix,
whereas the growth rate increases as the pitch becomes smaller in the experiment.
On the other hand, in agreement with the experiment we have observed the spontaneous generation of pairs of the polar-core vortices.

\section{Discussion}
\label{sec:discussion}

We here consider the possible reasons for the discrepancies discussed in the previous section.

\subsection{Initial noise dependence}
\label{sec:noise}

To simulate fluctuations and noises, we have introduced the initial noises of $\alpha, \beta, \gamma$ and $\delta$ 
in Eqs.~\eqref{eq:OP_pancake_initial} and \eqref{eq:initial}.
Here we calculate the spin dynamics for various variances of initial noises.

\begin{figure}[th]
\includegraphics[width=\linewidth]{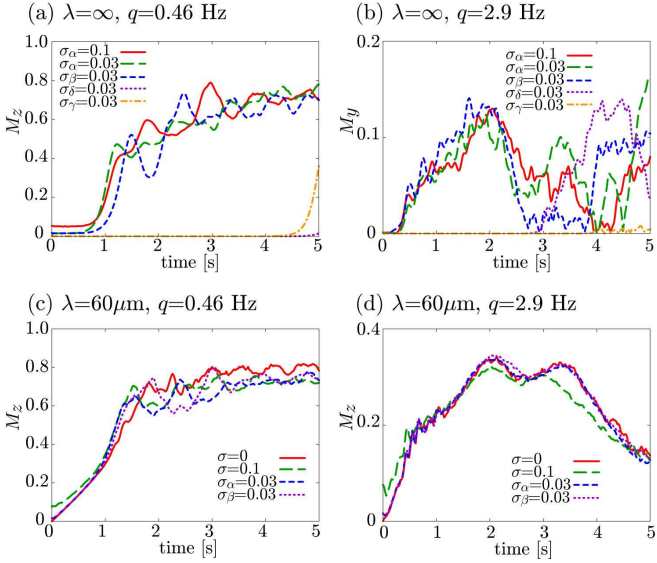}
\caption{
(Color online) Time evolution of spin fluctuations of
(a), (c), (d) $M_z\equiv \int d{\bm r} |f_z({\bm r})|/N$ 
and (b) $M_y\equiv |\int d{\bm r} f_y({\bm r})|/N$,
where $\sigma_\alpha=0.1$ means $\sigma_\alpha=0.1$ and other components vanish, etc.
In (c) and (d), $\sigma=0$ and 0.1 mean $\sigma_\alpha=\sigma_\beta=\sigma_\gamma=\sigma_\delta=0$ and 0.1, respectively.
}
\label{fig:noise}
\end{figure}

When we start with a uniform spin structure, 
the fluctuations in the transverse ($\alpha$) and longitudinal ($\beta$) magnetizations contribute to the growth of the instability,
while the fluctuations in the overall phase ($\gamma$) and amplitude of magnetization ($\delta$) hardly affect the formation of spin textures.
In Figs.~\ref{fig:noise} (a) and (b), we show the time evolution of the magnetization
starting from a uniform spin structure with various initial noises.
The fluctuation does not grow for $\sigma_\alpha=\sigma_\beta=0$ during a few seconds.
This result is consistent with the fact that
the unstable mode in an infinite quasi-2D system is proportional to $(1, 0, -1)^{\rm T}$ and decouples from
the fluctuations in the overall phase (phonon) and the amplitude of magnetization.
Due to the effect of the trapping potential,
the fluctuation eventually grows after a few seconds for $\sigma_\alpha=\sigma_\beta=0$.

On the other hand, when we start from a spin helix, 
the spin dynamics is independent of the detail of the initial noise,
since the spin current dominates the initial dynamics [Figs.~\ref{fig:noise} (c) and (d)].

\begin{figure}[th]
\includegraphics[width=0.7\linewidth]{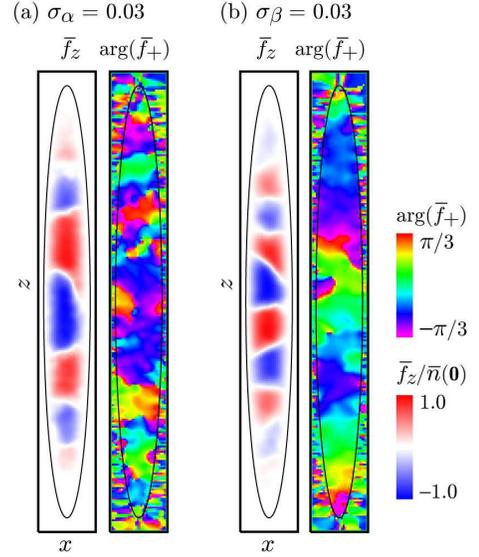}
\caption{
(Color) Magnetic structure at $t=5.3$~s starting from a uniform spin structure for $q/h=0.46$~Hz
with the initial noise of (a) $\sigma_\alpha=0.03$ and $\sigma_\beta=\sigma_\gamma=\sigma_\delta=0$ and
(b) $\sigma_\beta=0.03$ and $\sigma_\alpha=\sigma_\gamma=\sigma_\delta=0$.
The length scale of the fluctuation in the transverse magnetization is smaller (larger) than that in the longitudinal magnetization in (a) [(b)].
The size of each panel is $45~\mu{\rm m}\times 359~\mu{\rm m}$.
}
\label{fig:noise2}
\end{figure}

Although the magnetic pattern slightly depends on the initial noise,
the dependence is too little to account for the discrepancy between the experiment and the numerical result discussed in the previous section.
Figure~\ref{fig:noise2} shows an example of the initial noise dependence on the spin textures,
where the magnetic patterns at $t=5.3$~s
with the initial noise of (a) $\sigma_\alpha=0.03$ and $\sigma_\beta=\sigma_\gamma=\sigma_\delta=0$ and
(b) $\sigma_\beta=0.03$ and $\sigma_\alpha=\sigma_\gamma=\sigma_\delta=0$ are shown.
In both cases, $M_z$ saturates to $\sim 0.7$ and similar magnetic structures develop.
However, 
when the initial noise is transverse (longitudinal), 
the length scale of the spatial structure of the transverse (longitudinal) magnetization is smaller than the case when the initial noise is longitudinal (transverse),
which means 
the kinetic energy is stored in the fluctuation of the transverse (longitudinal) magnetization.

\subsection{Uncertainty of the parameters}
We examine the effect of the uncertainty of the coefficient $c_1$.
The strength of the spin exchange interaction has been measured
in molecular spectroscopy~\cite{Kempen2002} and in spin-mixing dynamics~\cite{Chang2005,Widera2006}.
According to these measurements, spin-exchange interaction energy $|c_1| n$ lies between 7 Hz and 13 Hz
for $n=2.8\times 10^{14}~{\rm cm}^{-3}$,
where the error bar in the difference of the scattering lengths is $-1.0a_{\rm B} < a_2-a_0 < -1.8 a_{\rm B}$.
A decrease in $|c_1|$ enhances the quadratic Zeeman effect.
Therefore, the effect of a nonuniform density distribution is enhanced
and the longitudinal magnetization induced by spin helix and the effect of the MDDI are suppressed.
However, the property of the magnetic pattern is qualitatively unchanged within the above range of the uncertainty
and it cannot resolve the discrepancy between the Berkeley experiment
and our calculation.

The uncertainty of the uniform magnetic field does not affect the spin dynamics
as long as the magnetic field is much stronger than the dipole field,
since we use the time-averaged dipole kernel~\eqref{eq:Q_rot_k} in the numerical simulations.
On the other hand, the residual field gradient may change the dynamics.
We calculate the spin dynamics under the residual field gradient $dB/dz=10~\mu{\rm G}/(2z_{\rm TF})$~\cite{Leslie_phD}.
However, the property of the magnetic pattern is almost unchanged in the short time scale of a few hundreds milliseconds,
although the residual field gradient winds helix more and more in the long time scale.

\subsection{Stable spin structure with the MDDI}
\label{sec:GS}
Next we consider the effect of the energy dissipation on the dynamics by replacing $t$ with $(1-i\Gamma)t$ in Eq.~\eqref{eq:3dGP}.
To keep the total longitudinal magnetization and number of atoms constant, 
we introduce the term $pm\Psi_m$ in the right-hand side of Eq.~\eqref{eq:3dGP} with $p$ being the Lagrange multiplier, 
and change $p$ and the chemical potential $\mu$ in each step.
The energy dissipation leads to enlargement of the spatial structure
rather than stabilization of the patterns, regardless of the value of $\Gamma$ over the range of $0.001 \le \Gamma \le 0.1$,
and eventually the system reaches a stationary state with a spatial structure of the order of 100~$\mu$m.

\begin{figure}[th]
\includegraphics[width=\linewidth]{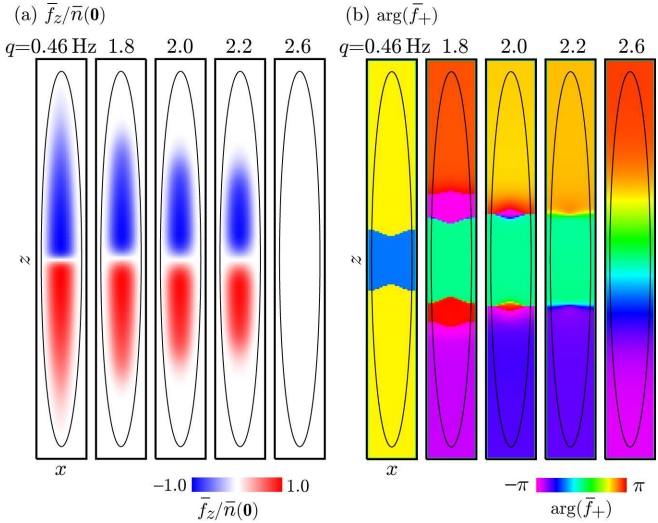}
\caption{(Color) Stationary configuration of (a) longitudinal magnetization $\bar{f}_z/\bar{n}_{\rm tot}({\bm 0})$
and (b) transverse magnetizations ${\rm arg}(\bar{f}_+)$ for the trap geometry and the number of atoms used in Ref.~\cite{Vengalattore2008}.
The condensate is almost fully magnetized in all cases (See Fig.~\ref{fig:stationary_M}).
The size of each panel is $45~\mu{\rm m}\times 359~\mu{\rm m}$.
}
\label{fig:stationary}
\end{figure}

\begin{figure}[th]
\includegraphics[width=0.9\linewidth]{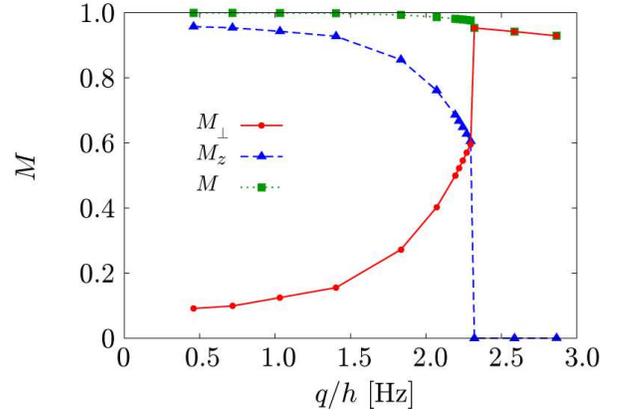}
\caption{(Color online) 
$q$ dependence of the amplitude of the transverse magnetization $M_\perp = \int d{\bm r} \sqrt{f_x^2+f_y^2}/N$,
longitudinal magnetization $M_z = \int d{\bm r} |f_z|/N$,
and total magnetization $M = \int d{\bm r} \sqrt{f_x^2+f_y^2+f_z^2}/N$, of the stationary state
for the trap geometry and the number of atoms used in Ref.~\cite{Vengalattore2008}.
}
\label{fig:stationary_M}
\end{figure}

We choose the same trap frequencies $(\omega_x,\omega_y,\omega_z)=2\pi\times (39, 440, 4.2)$~Hz and 
the number of atoms $N=2.3\times 10^6$ as used in Ref.~\cite{Vengalattore2008},
and investigate the stable spin configuration for various $q$.
The obtained results are shown in Fig.~\ref{fig:stationary},
and the $q$ dependence of the amplitude of the transverse, longitudinal, and total magnetizations are plotted in Fig.~\ref{fig:stationary_M}.
We have found two types of stable spin textures,
and the first-order phase transition between these structures occurs at $q=q_{\rm c}=h\times2.3$~Hz.
The critical quadratic Zeeman energy $q_{\rm c}$ is close to MDDI energy $E_{\rm dd}\simeq c_{\rm dd}n\tilde{\mathcal{Q}}_{\bm 0}$;
for the present system with $n=2.8\times 10^{14}{\rm cm}^{-3}$, $E_{\rm dd}\simeq h\times 1.9$~Hz.

The spin configuration is determined by the interplay between the MDDI and the quadratic Zeeman effect.
When the MDDI dominates the quadratic Zeeman energy,
the longitudinal magnetization is favored
since the spin-dependent factor $\delta_{\nu\nu'}-3\delta_{\nu z}\delta_{\nu' z}$ in Eq.~\eqref{eq:Q_rot} contributes maximally for $\nu=\nu'=z$.
From the orbital part in Eq.~\eqref{eq:Q_rot}, we find that the magnetic domain of the longitudinal magnetization elongates 
in the direction of the external magnetic field.
The condensate is almost fully magnetized and the magnetization at the domain wall is perpendicular to the magnetic field.
The direction of the domain wall is determined by the kinetic energy of the domain wall, and depends on the aspect ratio of the trap:
the domain wall is perpendicular to the magnetic field for the trap used in the Berkeley experiment whereas it is parallel to the magnetic field
in a pancake-shaped trap.

As the quadratic Zeeman energy becomes larger, the width of the domain wall becomes larger as shown in Fig.~\ref{fig:stationary} (a),
where the magnetization at the domain wall and at the top and bottom of the condensate is perpendicular to the magnetic field.
For $q>q_{\rm c}$, the quadratic Zeeman energy dominates the MDDI,
and magnetization occurs in the plane perpendicular to the magnetic field.
As regards transverse magnetization, the MDDI favors the antiferromagnetic ordering in the direction of the magnetic field,
resulting in the helical spin structure as shown for the case of $q/h=2.6$~Hz in Fig.~\ref{fig:stationary} (b).

The ground-state spin structure has also been investigated in Ref.~\cite{Kjall2009}
in a quasi-2D system by using the Metropolis Monte Carlo method.
Our results in Figs.~\ref{fig:stationary} and \ref{fig:stationary_M} are consistent with those in Ref.~\cite{Kjall2009}.

We have also searched a metastable state with a periodic pattern,
including the vortex lattice state of both MH and polar-core vortices suggested in Ref.~\cite{Zhang2009}.
We add a periodic modulation to the initial order parameter and investigate its dynamics with the imaginary-time propagation.
However, all configurations that we have prepared were unstable and eventually goes to the structures shown in Fig.~\ref{fig:stationary}.

\subsection{Spin vortex lattice}

\begin{figure}[th]
\includegraphics[width=\linewidth]{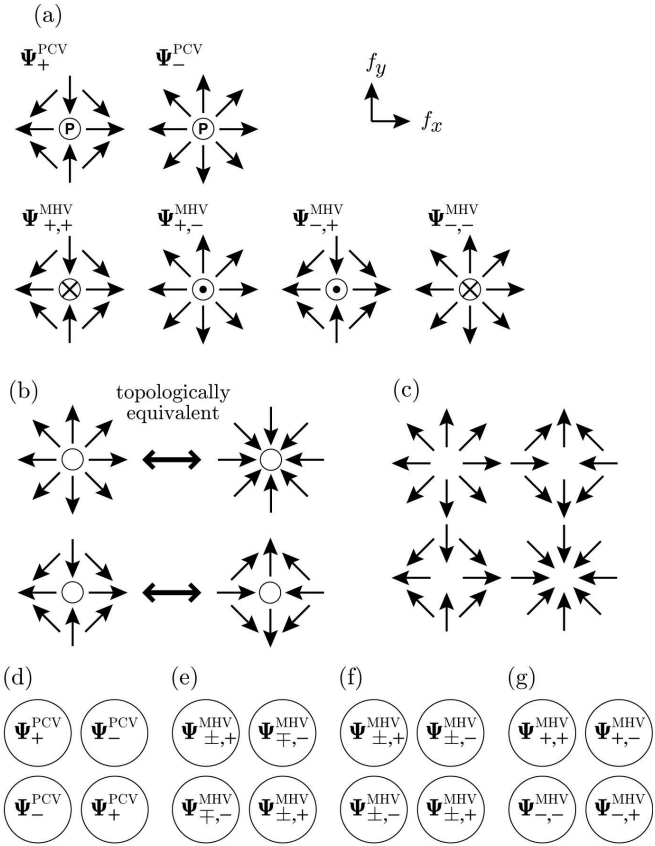}
\caption{(a) Spin configuration around vortices in Eqs.~\eqref{eq:pcv}--\eqref{eq:mhv},
where the arrows show the direction of the transverse magnetization,
{\sf P} indicates the polar core,
and $\bigodot$ and $\bigotimes$ mean that the magnetization at the vortex core points $+\hat{z}$ and $-\hat{z}$ direction, respectively.
(b) Two configurations are topologically equivalent if the core structure is the same.
(c) Unit of a vortex lattice that exhibits a periodic structure of the transverse magnetization.
(d)--(e) Possible sets of vortices to form a vortex lattice.
}
\label{fig:vortex_config}
\end{figure}

It is pointed in Ref.~\cite{Zhang2009} that
the vortex lattice of the MH vortices is long-lived in the imaginary-time propagation;
unfortunately we have not been able to reproduce such tendency.
To evaluate the lifetime of the vortex lattice,
we here investigate the real-time dynamics starting with the vortex-lattice state.
As mentioned before, there are two types of vortices in this system: the polar-core vortex and the MH vortex.
The order parameter around a polar-core vortex is given by
\begin{align}
 {\bm\Psi}^{\rm PCV}_{\pm}=\sqrt{n}\begin{pmatrix} e^{\pm i\varphi}f_v(r) \\ \sqrt{1-2f_v^2(r)}\\  e^{\mp i\varphi}f_v(r)\end{pmatrix},
\label{eq:pcv}
\end{align}
where we assume that the order parameter is axisymmetric around the vortex;
$\varphi$ is an azimuthal angle around the vortex,
and $f_v(r)$ is a monotonically increasing function which satisfies $f_v(0)=0$ and $\lim_{r\to \infty }f_v(r)=1/2$.
The spin current defined in Eq.~\eqref{eq:spincurrent} is anti-clockwise for ${\bm\Psi}^{\rm PCV}_+$ and
clockwise for ${\bm\Psi}^{\rm PCV}_-$, so that they form a vortex-antivortex pair.
The order parameter around a MH vortex is given by
\begin{align}
 {\bm\Psi}^{\rm MHV}_{+,\pm}&=\sqrt{n} e^{ i\varphi}
 \begin{pmatrix} e^{\pm i\varphi}\frac{1+\cos\beta_\pm(r)}{2} \\ \frac{\sin\beta_\pm(r)}{\sqrt{2}}\\  e^{\mp i\varphi}\frac{1-\cos\beta_\pm(r)}{2}
\end{pmatrix},\\
 {\bm\Psi}^{\rm MHV}_{-,\pm}&=\sqrt{n}e^{-i\varphi}
 \begin{pmatrix} e^{\pm i\varphi}\frac{1+\cos\beta_\mp(r)}{2} \\ \frac{\sin\beta_\mp(r)}{\sqrt{2}}\\  e^{\mp i\varphi}\frac{1-\cos\beta_\mp(r)}{2}
\end{pmatrix},
\label{eq:mhv}
\end{align}
where $\beta_+(r)$ and $\beta_-(r)$ are monotonically decreasing and increasing function, respectively,
which satisfy $\beta_+(0)=\pi$, $\beta_-(0)=0$ and $\lim_{r\to\infty}\beta_\pm(r)=\pi/2$.
The mass current
\begin{align}
{\bm j}^{\rm mass} &=\frac{\hbar}{2Mi}\sum_{m}[\Psi_m^*\nabla\Psi_{m} - (\nabla\Psi_m^*)\Psi_{m}]
\end{align}
is anti-clockwise for ${\bm\Psi}^{\rm MHV}_{+,\pm}$ and clockwise for ${\bm\Psi}^{\rm MHV}_{-,\pm}$,
while the spin current
is anti-clockwise for ${\bm\Psi}^{\rm MHV}_{\pm,+}$ and clockwise for ${\bm\Psi}^{\rm MHV}_{\pm,-}$.
Figure~\ref{fig:vortex_config} shows the spin structure around each vortex in Eqs.~\eqref{eq:pcv}--\eqref{eq:mhv} and
the possible configuration for the vortex lattice with periodic spin structure.
In order for the spin configuration to be periodic, 
the vortices with clockwise and anti-clockwise spin current 
have to align alternately as shown in Fig.~\ref{fig:vortex_config} (c).

\begin{figure}[th]
\includegraphics[width=\linewidth]{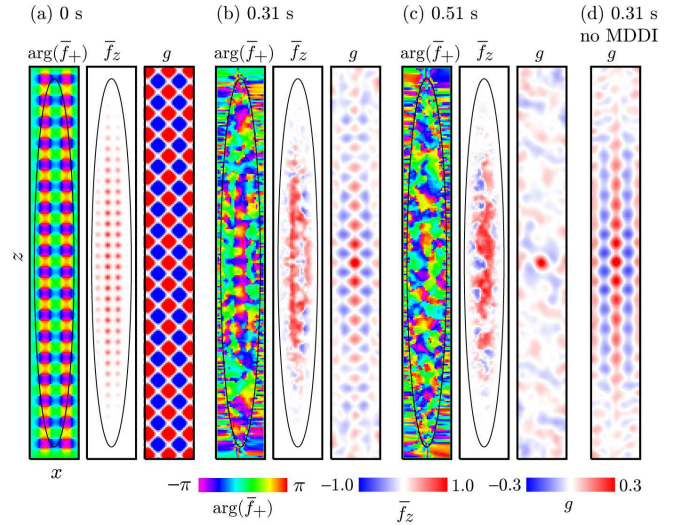}
\caption{(Color) Spin dynamics starting with spin-vortex lattice shown in Fig.~\ref{fig:vortex_config} (e) at $q/h=1.8$~Hz
for the trap geometry and the number of atoms used in Ref.~\cite{Vengalattore2008}. 
(a)--(c) Transverse magnetization ${\rm arg}(\bar{f}_+)$ (left), longitudinal magnetization $\bar{f}_z$ (middle), and spin correlation function $g$ (right)
calculated in the presence of the MDDI.
(d) Spin correlation function $g$ calculated in the absence of the MDDI.
Note that the periodic pattern in (d) is quite similar to that in the right panel of (b).
The size of each panel is $45~\mu{\rm m}\times 359~\mu{\rm m}$.
}
\label{fig:vortex}
\end{figure}

We prepare the spin-vortex-lattice state shown in Figs.~\ref{fig:vortex_config} (d)--(g)
with lattice constant $d_v$, and investigate the spin dynamics in the real-time evolution.
Here we use the ansatz $f_v(r)=\frac{1}{2}{\rm tanh}(r/\xi_{\rm sp})$,
$\beta_+(r)=\pi-\frac{\pi}{2}{\rm tanh}(r/\xi_{\rm sp})$, 
and $\beta_-(r)=\frac{\pi}{2}{\rm tanh}(r/\xi_{\rm sp})$.
Among the configurations in Figs.~\ref{fig:vortex_config} (d)--(g),
Fig.~\ref{fig:vortex_config} (f) is the most unstable and dissolves within 200~ms,
since this configuration has a nonzero mass circulation.
On the other hand, the periodic pattern survives longest
for the configuration shown in Fig.~\ref{fig:vortex_config} (e).
Figure~\ref{fig:vortex} shows the spin dynamics starting from the vortex lattice in Fig.~\ref{fig:vortex_config} (e)
with lattice constant $d_v=10~\mu$m.
In this case, the periodic pattern in the correlation function dissolves at around 0.5~s,
both in the presence and absence of the MDDI.

Our results show that
if the MH vortex-lattice emerges for some reason, it can survive for 0.5~s,
which is longer than the time scale for the emergence of the crystalline pattern in the experiment~\cite{Vengalattore2008}.
However, since spin dynamics in this system is slow, for instance, 
the helical spin configuration lasts more than 0.5~s as shown in Fig.~\ref{fig:B_cdd1_helix},
we cannot conclude that the spin vortex lattice is long-lived compared with other configurations.
Moreover, no mechanism has been presented
for the vortex lattice to appear in a short time scale of a hundred milliseconds.

\subsection{Thermalization due to the dipole relaxation}
\label{sec:dipole_relaxation}
Finally, we discuss the validity of the time-averaged dipole kernel~\eqref{eq:Q_rot_k}.
It is pointed in Ref.~\cite{Lamacraft2008} that the terms in the MDDI which are canceled by taking the time average induce the instability of the Larmor precession.
The wave number of the unstable mode is about $\sqrt{2M\omega_{\rm L}}/\hbar$.
This instability corresponds to the dipole relaxation, or the Einstein-de Haas effect~\cite{Kawaguchi2006a, Santos2006, Gawryluk2007}:
when two atoms in the highest Zeeman sublevel ($m=-1$) collide with each other, one or both of them can change their spin state to $m=0$ via the MDDI;
due to the energy conservation, the total kinetic energy of these atoms nearly equals to the linear Zeeman energy 
($\hbar\omega_{\rm L}$ or $2\hbar\omega_{\rm L}$ depending on the number of spin-flipped atoms).
The experimental result that the total longitudinal magnetization is conserved for a period longer than the time scale of the MDDI~\cite{Chang2004}
indicates that the effect of the dipole relaxation is quite small.
However, since
the linear Zeeman energy $\hbar\omega_{\rm L}/k_{\rm B}\simeq 5.5~\mu{\rm K}$ is much higher than
the temperature of the condensate ($\sim 100$~nK) in the present system,
a small number of spin-flipped atoms might thermalize the condensate.
The treatment of the thermal atoms is beyond the Gross-Pitaevskii formalism, and remains a challenge for the future work.

\section{Conclusions}
\label{sec:conclusions}
We have investigated the pattern formation dynamics in a spin-1 spinor dipolar Bose-Einstein condensate (BEC)
observed by the Berkeley group~\cite{Vengalattore2008},
by taking into account the effects of spinor and dipolar interactions,
linear and quadratic Zeeman energies, anisotropic trap geometries, various initial conditions, and noises.

We have first performed the Bogoliubov analysis in a uniform quasi-two-dimensional system,
and found that the spin helix does not enhance the instability but stabilize the system
in the presence of the magnetic dipole-dipole interaction (MDDI) (Figs.~\ref{fig:BdG_ene} and \ref{fig:BdG_mode}).
The minimum wavelength of the unstable mode is at least three times larger than 
the wavelength of the spin modulation observed in the experiment~\cite{Vengalattore2008}.
We have investigated the spin dynamics 
by simulating the three-dimensional Gross-Pitaevskii equation.
There are three mechanisms that generate spatial spin structures:
(i) a nonuniform density profile in an optical trap
induces checkerboard pattern,
even when we start from a uniform spin structure in the absence of the MDDI [Fig.~\ref{fig:B_cdd0} (a)];
(ii) the initial spin helix induces a spin current, which is reflected at the edge of the condensate
and generates a checkerboard pattern in the case for the trap geometry used in Ref.~\cite{Vengalattore2008} [Fig.~\ref{fig:B_cdd0} (b)],
and (iii) in agreement with the Bogoliubov analysis, the MDDI contributes to the pattern formation.
However, in all cases, the domain size of the obtained magnetic pattern ($\sim 15~\mu$m at minimum) is more than three times larger than
that observed in the experiment ($\lambda_{\rm exp}/2\sim 5~\mu$m).
It takes more than 500~ms for the pattern to develop in our calculation, whereas it develops within 200~ms in the experiment.

The other significant differences from the experiment are that
(i) when the magnetic pattern develops, it is always accompanied by the growth of the local longitudinal magnetization,
and (ii) the MDDI-induced dynamics strongly depends on the strength of the quadratic Zeeman energy $q$.

The detail of the initial noise does not qualitatively change the spin dynamics,
as long as it includes the fluctuations in transverse or longitudinal magnetizations.
We have also investigated the stationary spin structure in this system
and obtained different spin textures from that observed in the Berkeley experiment.
The stable texture undergoes the phase transition from a staggered domain of the longitudinal magnetization at small quadratic Zeeman energies
to a spin helix of the transverse magnetization at large quadratic Zeeman energies
due to the interplay between the quadratic Zeeman energy and the MDDI energy.
We have also considered the stability of the vortex-lattice state;
although the checkerboard pattern can survive for about 500~ms,
it is not long-lived compared with other spin textures.

From the above discrepancies, 
we conclude that the mean-field and Bogoliubov theories at zero temperature cannot account for the Berkeley experiment~\cite{Vengalattore2008}.
The effects absent in our calculation are many-body correlations and thermalization.
In particular, there might be a non-trivial effect of the thermal atoms via the dipole relaxation as discussed in Sec.~\ref{sec:dipole_relaxation}.
Since the system size in the direction of the strong confinement is smaller than the spin healing length,
the geometry of the condensate is two-dimensional with respect to the spin degrees of freedom.
Since the effects of fluctuations become prominent in low dimensional systems,
thermal fluctuations might contribute significantly to the magnetic pattern.
If experimental external noises are not the origin of the magnetic pattern,
it would be a new quantum phenomenon beyond the mean-field theory.
The clarification of these issues remains a challenge for a future work.

\begin{acknowledgments}
YK and MU thank Mukund Vengalattore for valuable discussions.
This work was supported by MEXT (KAKENHI 20540388, 22340114, 22340116, and 22740265,
the Global COE Program ``the Physical Sciences Frontier'', and the Photon Frontier Network Program),
and JSPS and FRST under the Japan-New Zealand Research Cooperative Program.
\end{acknowledgments}

%\newpage %Just because of unusual number of tables stacked at end
%\bibliography{References}% Produces the bibliography via BibTeX.

%{\it Note added.}
%After completion of the present work, a preprint~\cite{Zhang2009} and a paper~\cite{Kjall2009} appeared
%which also discusses stationary magnetic patterns in a spinor dipolar BEC.
%The stable spin structure calculated using Monte Calro simulation in Ref.~\cite{Kjall2009} is consistent with our result.

%\bibliography{dipole.bib}% Produces the bibliography via BibTeX.
%\begin{thebibliography}{99}

%\end{thebibliography}

\end{document}